\documentclass[11pt,letterpaper]{article}
\usepackage{amsmath}
\usepackage{amsfonts}
\usepackage{amssymb}
\usepackage{graphicx}
\usepackage[utf8]{inputenc}
\usepackage[english]{babel}

\usepackage{fullpage}
\usepackage{lscape}
\usepackage{hyperref}
\usepackage{textcomp}
\usepackage{bbold}
\usepackage{diagbox}
\usepackage{xcolor}
\usepackage{pdfpages}
\usepackage{indentfirst}
\usepackage{cancel}
\usepackage{tensor}
\usepackage{slashed}
\usepackage[titletoc]{appendix}
\usepackage[left=2cm,right=2cm,top=2cm,bottom=2cm]{geometry}
\usepackage{authblk}
\hypersetup{
	colorlinks=true,
    linkcolor={black},
    citecolor={blue!80!black},
    urlcolor={magenta!60!black},
}

\newcommand*{\bfrac}[2]{\genfrac{\lbrace}{\rbrace}{0pt}{}{#1}{#2}}

\title{Einstein-Cartan-Dirac equations in the Newman-Penrose formalism}

\author[1]{Swanand Khanapurkar}
\author[2]{Abhinav Varma}
\author[3]{Nehal Mittal}
\author[4]{Navya Gupta}
\author[5]{Tejinder P. Singh}
\affil[1]{\textit{Indian Institute of Science Education and Research (IISER), Pune 411008, India.}}
\affil[2]{\textit{University College London (UCL), London WC1E 6BT, United Kingdom.}} 
\affil[3]{\textit{Indian Institute of Technology Bombay (IITB), Mumbai 400076, India.}}
\affil[4]{\textit{Indian Institute of Technology Kanpur (IITK), Kanpur 208016, India.}}
\affil[5]{\textit{Tata Institute of Fundamental Research (TIFR), Mumbai 400005, India.}}

\date{}

\begin{document}
	
\maketitle

\texttt{E-mail: $^1$\url{swanand.khanapurkar@students.iiserpune.ac.in}, $^2$\url{abhinav.varma.17@ucl.ac.uk}, \\ $^3$\url{14D260006@iitb.ac.in},$^4$\url{navyag@iitk.ac.in},$^5$\url{tpsingh@tifr.res.in}}
\bigskip

\begin{abstract}
\noindent We formulate the Einstein-Cartan-Dirac equations in the Newman-Penrose (NP) formalism, thereby presenting a more accurate and explicit analysis of previous such studies. The equations show in a transparent way how the Einstein-Dirac equations are modified by the inclusion of torsion. In particular, the Hehl-Datta equation is presented in NP notation. We then
describe a few solutions of the Hehl-Datta equation on Minkowski space-time, and in particular report a solitonic solution which removes the unphysical behavioiur of the corresponding Dirac solution. The present work serves as a prelude to similar studies for non-degenerate Poincar\'e gauge gravity.

\end{abstract}

\tableofcontents

\section*{Notation and conventions}
The following conventions are in use for the remainder of this paper:

\begin{itemize}

\item[$\square$] The Lorentz Signature used is (+ - - -) throughout.

\item[$\square$] $V_4$ is a non-torsional space-time, while a space-time endowed with torsion is specified by $U_4$.


\item[$\square$] Greek indices, e.g. $\alpha, \zeta, \delta $ refer to world components, which transform according to \textit{general coordinate transformations} and are raised or lowered using the metric $g_{\mu\nu}$.

\item[$\square$] Latin indices within parenthesis e.g. (a) or (i) are tetrad indices, which transform according to \textit{local Lorentz transformations} in the flat tangent space, and are raised or lowered using $\eta_{(i)(k)}$.

\item[$\square$] Latin indices (without parenthesis) e.g. $i,j,b,c$ indicate objects in Minkowski space, which transform according to \textit{global Lorentz transformations}. 

\item[$\square$] In general $0,1,2,3$ refer to world indices while $(0),(1),(2),(3)$ refer to tetrad indices.

\item[$\square$] The total covariant derivative is denoted by $\nabla$, while $\{\}$ denotes the Christoffel connection. Correspondingly, $\nabla^{\{\}}$ represents a covariant derivative with respect to the Christofell connections.

\item[$\square$] Commas $(,)$ indicate partial derivatives while semicolons $(;)$ indicate the Riemannian covariant derivative. Thus, for tensors, $;$ and $\nabla^{\{\}}$ are same, while for spinors, $(;)$ involves both partial derivatives and the Riemannian part of the spin connection, $\gamma$, as defined in the following.

\item[$\square$] The four component Dirac-spinor is written as
	\begin{equation}\label{diracspinor} 
	\psi = \begin{bmatrix} P^A \\ \bar{Q}_{B'} \end{bmatrix}  
	\end{equation}
	where $P^A$ and $\bar{Q}_{B'}$ are two dimensional complex vectors in $\mathbb{C}^2$ space. We redefine the spinors as: $P^0 = F_1$, $P^1 = F_2$, $\bar{Q}^{1'} = G_1$ and $\bar{Q}^{0'} = -G_2$. This is in accordance with our primary source, \cite{Chandru}, the notations, conventions and representations wherein are generally adhered to in this paper.

\end{itemize}

\newpage
\section{Introduction}
Einstein's general theory of relativity (GR), published in 1915, has been described as the most beautiful of all the existing physical theories \cite{landau_2}. The background space-time on which classical GR is formulated is a Riemannian manifold (denoted $V_4$) which is torsion-less. In this case, the affine connection coincides uniquely with the Levi-Civita connection and geodesics coincide with the path of shortest distance. This is, however, not generally true for other, \textit{torsional} manifolds, such as the manifold on which the Einstein-Cartan-Sciama-Kibble (ECSK) -- or simply, Einstein-Cartan (EC) -- theory is formulated. In such a theory, the geometrical structure of the manifold is modified such that the affine connection is no longer required to be symmetric, and no longer coincides uniquely with the Levi-Civita connection
t\cite{Cartan_1922, Hehl1971, hehl_RMP, sciama_GR, sciama_physica_structure_of_GR, kibble_GR}. 

Torsion, as an antisymmetric part of the affine connection was introduced by Elie-Cartan (1922) \cite{Cartan_1922}. Also termed the $U_4$ theories of gravitation, Einstein-Cartan theories work with an underlying manifold that is non-Riemannian (unlike classical GR which is formulated on $V_4$). The non-Riemannian part of the manifold is associated with the spin density of matter, which plays the role of a source analogous to the role of mass in Riemannian curvature. Here, mass and spin \textit{both} play the dynamical role. While mass ``adds up" on classical length scales due to its monopole character, spin, being of dipole character, usually averages out in the absence of external forces.

For this reason, matter, in the macro-physical regime, can be dynamically characterized entirely by the energy-momentum tensor. In the micro-regime, heuristic arguments suggest that a spin density tensor plays an analogous role for spin, and related, as with mass and curvature, to some other geometrical property of space-time. It is this requirement that EC/ECSK theory satisfies (the reader is referred to \cite{hehl_RMP} for a detailed treatment). When we minimally couple the Dirac field on $U_4$, we term this \textit{Einstein-Cartan-Dirac (ECD) theory}. There are two independent geometric fields -- the metric and torsion -- and one matter field $\psi$ in this theory. Varying the corresponding Lagrangian, we get three equations of motion, corresponding to the modified Einstein field equations, modified Dirac equation, and a torsional coupling. On $U_4$, the Dirac equation on $U_4$ becomes non-linear; and is known as  the \textit{Hehl-Datta (HD) equation} after the seminal work in \cite{Hehl1971}. 

The usual method in approaching solutions to problems in GR is to use a \textit{local coordinate basis} $\hat{e}^{\mu}$ such that $\hat{e}^{\mu} = {\partial_{\mu}}$. This coordinate basis field is covariant under general coordinate transformations. However, it has been found useful to employ non-coordinate basis techniques in problems involving spinors. Moreover, choosing the tetrad basis vectors as \textit{null vectors} is extremely useful in some situations. This formalism, where a given theory is expressed in a basis of null tetrads, is the celebrated Newman-Penrose (NP) formalism. In this formalism, we replace tensors by their null tetrad components and represent these components with certain distinctive symbols. Most of the important and physically relevant geometrical objects and identities (e.g. the Riemann curvature tensor, Weyl tensor, Bianchi identities, Ricci identities etc.) on $U_4$ have been formulated in the NP formalism (such as in \cite{jogia_Griffiths}). 

It can be shown that there is a natural connection between spin dyads (a detailed account of spin dyads can be found in \cite{Chandru}) and null tetrads \cite{Chandru,SVD_geometry_fields_cosmology}. Physical systems involving spinor fields can be fully expressed in the NP formalism (for example, the Dirac equation on $V_4$ has been studied extensively, ref. Chapter 12 in \cite{Chandru}). In addition, many systems in gravitational physics are also studied in the NP formalism \cite{Chandru}. It appears that the NP formalism is the shared vocabulary between the physics of relativistic quantum mechanical systems (with spinor fields) and classical gravitational systems (having a metric and/or torsion). 

In the present paper, we aim to formulate the full ECD equations in the NP formalism. We know that the contorsion tensor is completely expressible in terms of the Dirac state \cite{hehl_RMP}. We wish to then find expressions for the contorsion spin coefficients -- which are the standard NP variables that account for spin -- explicitly in terms of the Dirac state. Using this, we can write the complete set of HD equations in the NP formalism. In a sense, this work is to be read as a sequel to the work of S. Chandrasekhar in (Chapter 12 of) \cite{Chandru}, where Dirac equation in $V_4$ has been given a full treatment in the NP formalism. Some recent works {\cite{Timofeev2016,zecca_NP,zecca_NP1} attempt to do that but have not provided explicit corrections to the standard NP variables due to torsion. Further, there are notational and sign inconsistencies in many such examples of existing literature in the field, and we aim to provide a comprehensive and self-contained treatment.

Finally, we attempt at solutions to the HD equations in a Minkowski space with torsion. This, apart from being the simplest case to consider, is also motivated by certain physical intuitions which can be considered as supporting, but non-essential, corollaries to this work. A recent essay, \cite{TP_1,TP_2,GRFessay2018}, suggests the incorporation of a new length scale in quantum gravity, thereby providing a symmetry between large and small masses; a conjecture has been proposed therein to establish a duality between these two limits. This conjecture is predicated on the necessary existence of solutions to the Hehl-Datta equations on Minkowski space, representing the balance between the Riemannian and torsional effects which reduce to small and large masses in the respective limits. However, notwithstanding the duality conjecture and the new length scale proposed, our results hold for the standard theory as well. All equations are expressed in terms of two relevant generic length scales, $l_1 = L_{Pl}$ and $l_2 = \frac{1}{2}\lambda_C$, the first being Planck length, and the second being one half of the Compton wavelength. In case of the modified ECD theory with a new length scale $L_{CS}$ (as defined below), we will instead have
$l_1=l_2=L_{CS}$: Planck length and Compton wavelength no longer appear in the ECD equations, and are both replaced 
by $L_{CS}$.

\section{Einstein-Cartan theory and its coupling to the Dirac field}

\subsection{Einstein-Cartan theory}

In the Einstein-Cartan theory, the Riemannian manifold of ordinary GR ($V_4$) is promoted to the corresponding non-Riemannian manifold $U_4$. As discussed, this latter manifold admits, in addition to the structure of ordinary GR, a non-vanishing torsion. Torsion is a (rank 3) tensorial object defined as the antisymmetric part of the affine connection: 
\begin{equation}\label{eq:torsiontensor}
Q\indices{_{\alpha\beta}^\mu} = \Gamma\indices{_{[\alpha\beta]}^\mu} = \frac{1}{2}(\Gamma\indices{_{\alpha\beta}^\mu} - \Gamma\indices{_{\beta\alpha}^\mu})
\end{equation} 
\indent Similarly, the \textit{contorsion} tensor $K\indices{_{\alpha\beta}^\mu}$ is given by $K\indices{_{\alpha\beta}^\mu} = -Q\indices{_{\alpha\beta}^\mu} - Q\indices{^\mu_{\alpha\beta}} + Q\indices{_\beta^\mu_\alpha}$. This allows us to write -- in terms of the usual Christoffel symbols -- the following relation:
\begin{equation}\label{eq:affine connection}
\Gamma\indices{_{\alpha\beta}^\mu} = \bfrac{\mu}{\alpha\beta}- K\indices{_{\alpha\beta}^\mu}
\end{equation} 
\indent When a matter field $\psi$ is minimally coupled with gravity and torsion,  its action is given as follows \cite{hehl_RMP}:
\begin{equation} \label{ECmatteraction}
S = \int d^{4}x \sqrt{-g} \Big[\mathcal{L}_{m} (\psi, \nabla\psi, g) - \frac{1}{2k} R(g, \partial g)\Big]
\end{equation}
\indent Here $k = {8\pi G}/{c^4}$, $\mathcal{L}_{m}$ is the matter Lagrangian density, and the second term represents the Lagrangian density for the gravitational field. There are three fields in this Lagrangian: $\psi$, $g_{\mu\nu}$, and $K_{\alpha\beta\mu}$, representing the matter field, the metric, and the contorsion, respectively. Varying the action with respect to these, one arrives at the following three field equations:
\begin{align}
\frac{\delta(\sqrt{-g}\mathcal{L}_m)}{\delta \psi} &= 0  \label{eq:generic EOM of matter}\\
\frac{\delta(\sqrt{-g}R)}{\delta g^{\mu\nu}} &= 2k \frac{\delta(\sqrt{-g}\mathcal{L}_{m})}{\delta g^{\mu\nu}} \label{eq:generic EOM of metric-mass}\\
\frac{\delta(\sqrt{-g}R)}{\delta K_{\alpha\beta\mu}} &= 2k \frac{\delta(\sqrt{-g}\mathcal{L}_{m})}{\delta K_{\alpha\beta\mu}} \label{eq:generic EOM of torsion-spin}
\end{align}
\indent Here, (\ref{eq:generic EOM of matter}) leads us to the matter field equations on a curved space-time with torsion. The right hand side of (\ref{eq:generic EOM of metric-mass}) is associated with $\sqrt{-g}k T_{\mu\nu}$ via the definition of $T_{\mu\nu}$, the metric energy-momentum tensor. Similarly, the right hand side  of (\ref{eq:generic EOM of torsion-spin}) is associated with $2\sqrt{-g}k S^{\mu\beta\alpha}$ where $S^{\mu\beta\alpha}$ is the spin density tensor. Together, these two yield the Einstein-Cartan field equations:
\begin{equation}
G^{\mu\nu} = k \Sigma^{\mu\nu}
\label{TPS1}
\end{equation}
\begin{equation}
T^{\mu\beta\alpha} =k S^{\mu\beta\alpha}
 \label{TPS2}
 \end{equation}
\indent In (\ref{TPS1}) the $G^{\mu\nu}$ on the left hand side is the asymmetric Einstein tensor built from the asymmetric connection, while $\Sigma^{\mu\nu}$ is the asymmetric canonical (total) energy momentum tensor, constructed out of the symmetric (metric) energy-momentum tensor and the spin density tensor. In (\ref{TPS2}), the so-called `modified' torsion $T^{\mu\beta\alpha}$ is the traceless part of the torsion tensor, and is algebraically related to $S^{\mu\beta\alpha}$ on the right. In the limit torsion $\rightarrow 0$, we recover classical GR -- (\ref{TPS2}) vanishes, and (\ref{TPS1}) reduces to the Einstein field equations which couple the (symmetric) Einstein tensor to the (symmetric) metric energy-momentum tensor.

\subsection{EC coupling to the Dirac field}

The theory generated from the minimal coupling of the Dirac field on $U_4$ is what we term \textit{Einstein-Cartan-Dirac (ECD) theory}. In this theory, the matter field is the spinorial Dirac field $\psi$, for which the Lagrangian density is given by (note the noncommuting covariant derivatives):
\begin{equation}\label{diraclagrangian}
\mathcal{L}_{m} = \frac{i\hbar c}{2}(\overline{\psi}\gamma^{\mu}\nabla_{\mu}\psi - \nabla_{\mu}\overline{\psi}\gamma^{\mu}\psi) - mc^{2}\overline{\psi}\psi
\end{equation}

In ECD theory, the addition of spin degrees of freedom necessitates a more careful treatment of anholonomic objects.  As we define the affine connection, $\Gamma$, to facilitate parallel transport of geometrical objects with world (Greek) indices, so do we define the spin connection $\gamma$ for anholonomic objects (with Latin indices). The affine connection can be decomposed into a Riemannian ($\{\}$) and a torsional part (made up of the contorsion tensor, $K$) and similarly, the spin connection $\gamma$ can also be decomposed into a Riemannian ($\gamma^o$) and torsional part (once again, formed of the contorsion tensor). These components are related via the following equation (following the notation in \cite{jogia_Griffiths}):  
\begin{equation}\label{eq:spinconnection}
\gamma\indices{_\mu^{(i)(k)}} = 
\tensor*{\gamma}{_\mu^o}\indices{^{(i)(k)}} - 
K\indices{_\mu^{(k)(i)}}
\end{equation}
where $\tensor*{\gamma}{_\mu^o}\indices{^{(i)(k)}}$ is Riemannian part and $K\indices{_\mu^{(k)(i)}}$ is the torsional part.
Using these, we define the covariant derivative for spinors, on $V_4$ and $U_4$:
\begin{align}
&\psi_{;\mu} = \partial_{\mu} \psi + \frac{1}{4}\gamma^o_{\mu (b)(c)}\gamma^{[(b)}\gamma^{(c)]}\psi \: \: \: \: \: \: \: \: \: \: \: \: \: \: \: \: \: \: \: \: \: \: \: \: \: \: \: \: \: \: \: \: \: \: \: \: \: \: \: \: \: \: \: \: \: \: \: \: \: \: \: \: \: \: \: \: (\text{on} \: \: V_4) \label{covariantderivativev4}\\
&\nabla_{\mu} \psi = \partial_{\mu} \psi +  \frac{1}{4}\gamma^0_{\mu (c)(b)}\gamma^{[(b)}\gamma^{(c)]}\psi - \frac{1}{4} K_{\mu (c)(b)}\gamma^{[(b)}\gamma^{(c)]}\psi \: \: \: \: \: \: \: \: \: \: \: (\text{on} \: \: U_4) \label{covariantderivative}
\end{align}
\indent Substituting this into (\ref{diraclagrangian}) we obtain the explicit form of Lagrangian density; varying with respect to $\bar{\psi}$ as in (\ref{eq:generic EOM of matter}) 
yields the Dirac equation on $V_4$ and $U_4$:

\begin{align}
&i\gamma^{\mu}\psi_{;\mu}- \frac{mc}{\hbar}\psi = 0 \: \: \: \: \: \: \: \: \: \: \: \: \: \: \: \: \: \: \: \: \: \: \: \: \: \: \: \: \: \: \: \: \: \: \: \: \: \: \: \: \: \: \: \: \: \: \: \: \: \: \: \: \: \: \: \: \: \: \: \: \: \: \: \: \: \: \: \: \: \: \: \: \: \: \: \: \: \: \: \: \: (\text{on} \: \: V_4) \label{eq:diracv4}\\
&i\gamma^{\mu}\psi_{;\mu} + \frac{i}{4}K_{(a)(b)(c)}\gamma^{[(a)}\gamma^{(b)}\gamma^{(c)]}\psi - \frac{mc}{\hbar}\psi = 0  \: \: \: \: \: \: \: \: \: \: \: \: \: \: \: \: \: \: \: \: \: \: \: \: \: \: \: \: \: (\text{on} \: \: U_4)\label{eq:diracu4}
\end{align}
\indent Next, we use (\ref{eq:generic EOM of metric-mass})
and Lagrangian density defined in (\ref{diraclagrangian}) to obtain the gravitational field equations on $V_4$ and $U_4$:

\begin{align}
&G_{\mu\nu}(\{\}) = \frac{8\pi G}{c^4} T_{\mu\nu} \: \: \: \: \: \: \: \: \: \: \: \: \: \: \: \: \: \: \: \: \: \: \: \: \: \: \: \: \: \: \: \: \: \: \: \: \: \: \: \: \: \: \: \: \: \: \: \: \: \: \: \: \: \: \: \: \: \: \: \: \: \: \: \: \: \: \: \: \: \: \: \: \: \: \: \: \: \: \: \: (\text{on} \: \: V_4)    \label{eq:efev4}\\
&G_{\mu\nu}(\{\}) = \frac{8\pi G}{c^4} T_{\mu\nu} - \frac{1}{2} \bigg{(}\frac{8\pi G}{c^4}\bigg{)}^2 g_{\mu\nu}  S^{\alpha\beta\lambda}S_{\alpha\beta\lambda} \: \: \: \: \: \: \: \:\: \: \: \: \: \: \: \: \: \:\: \: \: \: \: \: \: \: \: \: (\text{on} \: \: U_4)   \label{eq:efeu4}
\end{align}
\indent Here, $T_{\mu\nu}$ is the metric EM tensor which is symmetric and defined as:
\begin{equation}\label{dynamic EM tensor}
T_{\mu\nu} = \Sigma_{(\mu\nu)}(\{\}) = \frac{i\hbar c}{4}\Big[\bar{\psi}\gamma_{\mu} \psi_{;\nu} + \bar{\psi}\gamma_{\nu} \psi_{;\mu} - \bar{\psi}_{;\mu} \gamma_{\nu}\psi -\bar{\psi}_{;\nu} \gamma_{\mu}\psi  \Big]
\end{equation}
\indent Equations (\ref{eq:diracv4}) and (\ref{eq:efev4}) together form the system of equations of Einstein-Dirac theory.
We now move to the full Einstein-Cartan-Dirac theory. Using the Lagrangian density defined in (\ref{diraclagrangian}), we can define the spin density tensor:

\begin{equation}\label{eq:spindensity}
S^{\mu\nu\alpha} = \frac{-i\hbar c}{4}\bar{\psi}\gamma^{[\mu}\gamma^{\nu}\gamma^{\alpha]}\psi
\end{equation}
\indent Using (\ref{eq:spindensity}) and (\ref{eq:generic EOM of torsion-spin}),
(\ref{eq:diracu4}) can be simplified to give the Hehl-Datta equation \cite{hehl_RMP}, \cite{Hehl1971}. This, along with (\ref{eq:efeu4}) and the relation between the modified torsion tensor and spin density tensor, define the field equations of the {Einstein-Cartan-Dirac theory: 
	\begin{align}
	G_{\mu\nu}(\{\}) &= \frac{8\pi G}{c^4} T_{\mu\nu} - \frac{1}{2} \bigg{(}\frac{8\pi G}{c^4}\bigg{)}^2 g_{\mu\nu}  S^{\alpha\beta\lambda}S_{\alpha\beta\lambda}\label{eq:ECDgravity}\\
	T_{\mu\nu\alpha} &=  - K_{\mu\nu\alpha} = \frac{8\pi G}{c^4}S_{\mu\nu\alpha} \label{eq:ECDtorsion}\\
	i\gamma^{\mu}\psi_{;\mu} &= +\frac{3}{8}L_{Pl}^{2}\overline{\psi}\gamma^{5}\gamma_{(a)}\psi\gamma^{5}\gamma^{(a)}\psi + \frac{mc}{\hbar}\psi   \label{eq:HD}\\
	\end{align}  
	where $L_{Pl}$ is the Planck length.

\section{\texorpdfstring{Introducing a unified length scale $L_{CS}$ in quantum gravity }%
{Introducing a unified length scale LCS in quantum gravity}}\label{sec:Introduction to LCS}

Recent work \cite{TP_1}, \cite{TP_2} has provided motivation for unifying the Compton wavelength ($\frac{\lambda}{\hbar c}$) and Schwarzschild radius ($R_s = \frac{2GM}{c^2}$) of a point particle with mass $m$ into one single length scale, the Compton-Schwarzschild length ($L_{CS}$). Such a treatment compels us to introduce torsion, and relate the Dirac field to the torsion field. An action principle has been proposed with this new length scale which permits the Dirac equation and the Einstein field equations as mutually dual limiting cases. The modified action proposed is as follows: 
\begin{equation}
\dfrac{L_{Pl}^2}{\hbar} S = \int d^4x \sqrt{-g} \: \Big[R-\frac{1}{2} L_{CS}\bar{\psi}\psi +  L_{CS}^2 \bar{\psi} i \gamma^{\mu} \nabla_{\mu}\psi \Big]
\end{equation} 
\indent Using this new length scale, $L_{CS}$, we can rewrite the Einstein-Cartan-Dirac equations as \cite{TP_2}:
\begin{align}
G_{\mu\nu}(\{\}) &= \frac{8\pi L_{CS}^{2}}{\hbar c} T_{\mu\nu} + \left(\frac{8\pi L_{CS}^{2} }{\hbar c}\right)^2 \tau_{\mu\nu}\label{ECDLCSgravity}\\
T_{\mu\nu\gamma} &= \frac{8\pi L_{CS}^{2}}{\hbar c} S_{\mu\nu\gamma} \label{ECDLCStorsion}\\
i\gamma^a \psi_{;a} &= +\frac{3}{8} L_{CS}^2 \bar{\psi}\gamma^5\gamma_{a}\psi\gamma^5\gamma^a\psi + \frac{1}{2L_{CS}} \psi = 0 \label{HDLCS}
\end{align}
\indent A note on length scales: We use $l$ to denote a length scale in the theory. For standard ECD theory, the two scales that appear  are the Planck length $l_1 = L_{Pl} = \sqrt{\frac{G\hbar}{c^3}}$, and half the Compton wavelength $l_2 = \frac{\lambda_C}{2} = \frac{\hbar}{2mc}$. For the modified ECD theory, we have $l_1 = l_2 = L_{CS}$, in terms of the new unified length scale.

\section{The Newman-Penrose formalism and ECD in NP}

\subsection{Tetrads}
It is common in the literature \cite{Chandru} to use tetrads (or \textit{vierbeins}) to define spinors on a curved space-time (in $V_4$ as well as $U_4$)\footnote{While this is often the case, there are other formalisms that can be used \cite{weldon}}. In this formalism, the transformation properties of spinors are defined in a flat (Minkowski) space, locally tangent to $U_4$. We know that at each point in space-time, we can define a coordinate basis vector field $\hat{e}^{\mu} = {\partial_{\mu}}$ which is covariant under general coordinate transformations. The basis vectors associated with spinors, however, are covariant under \textit{local} Lorentz transformations. Hence, we define, at each point of our manifold, a set of four orthonormal basis vectors (forming the tetrad field) given by $\hat{e}^i (x)$. These comprise four vectors (one for each $\mu$) at each point, and the tetrad field is governed by the relation ${\hat{e}^i (x)} = e^i_{\mu} (x) \hat{e}^{\mu}$ where the transformation matrix $e^i_{\mu}$ is such that:
\begin{equation}\label{eq:tetradmetrictransformation}
e^{(i)}_{\mu} e^{(k)}_{\nu} \eta_{{(i)}{(k)}} = g_{\mu\nu}
\end{equation}
\indent The transformation matrix $e^{(i)}_\mu$ allows us to convert the components of any world tensor (a tensor which transforms according to general coordinate transformations) to the corresponding components in a local Minkowskian space (the latter of these being covariant under local Lorentz transformations). 
\subsection{Introduction to the NP formalism}
The Newman-Penrose (NP) formalism was formulated by Newman and Penrose in their work \cite{NP_original_paper}. It is a special case of tetrad formalism; where we choose our tetrad as a set of four null vectors: 
\begin{equation}
e_{(0)}^{\mu} = l^{\mu},~~~ e_{(1)}^{\mu} = n^{\mu},~~~ e_{(2)}^{\mu} = m^{\mu}, ~~~e_{(3)}^{\mu} = \bar{m}^{\mu}
\end{equation}
where $l^{\mu}, n^{\mu}$ are real and $m^{\mu}, \bar{m}^{\mu}$ are complex. The null tetrad indices are raised and lowered using the flat space-time metric 
\begin{equation}
\eta_{(i)(j)} = \eta^{(i)(j)} = \begin{pmatrix}
0 & 1& 0 & 0 \\
1 & 0& 0 & 0 \\
0 & 0& 0 & -1 \\
0 & 0& -1 & 0 \\
\end{pmatrix}
\end{equation}
and the tetrad vectors satisfy the equation $g_{\mu\nu} = e_{\mu}^{(i)}e_{\nu}^{(j)}\eta_{(i)(j)} $. In this formalism, we replace tensors by their tetrad components and represent these components with a collection of distinctive symbols which are now standard in the literature.

\subsection{Spinor analysis}

We define four null tetrads (and their corresponding co-vectors) on Minkowski space (raised and lowered using $\eta_{\mu\nu}$):
\begin{equation}
l^a = \frac{1}{\sqrt{2}} (1,0,0,1), ~ m^a =  \frac{1}{\sqrt{2}} (0,1,-i,0),~ \bar{m}^a = \frac{1}{\sqrt{2}} (0,1,i,0), ~ n^a = \frac{1}{\sqrt{2}} (1,0,0,-1)  \\
\end{equation}
\indent We also define the following Van der Waarden symbols:
\begin{equation}\label{vanderwaarden}
	\sigma^a = \sqrt{2}\begin{bmatrix} l^a & m^a \\ \bar{m}^a & n^a \end{bmatrix}~~~~~~~~~~~~~~~~~~~~
	\tilde{\sigma}^a = \sqrt{2}\begin{bmatrix} n^a & -{m}^a \\ -\bar{m}^a & l^a \end{bmatrix}  
	\end{equation}
\indent For the Dirac gamma matrices, we use the complex version of the Weyl (chiral) representation:
\begin{equation}\label{gammamatrices}
	\gamma^a = \begin{bmatrix} 0 & (\tilde{\sigma}^a)^* \\ (\sigma^a)^* & 0 \end{bmatrix} ~~~\text{where}~~~ \gamma^0 = \begin{bmatrix} 0 & \mathbb{1} \\ \mathbb{1} & 0 \end{bmatrix},~~ \gamma^i = \begin{bmatrix} 0 & (-\sigma^{i})^* \\ (\sigma^{i})^* & 0 \end{bmatrix}
	\end{equation}
where $a = (0,1,2,3)$.

The complex Weyl representation is used so that the Dirac bispinor and gamma matrices defined in (\ref{diracspinor}) and (\ref{gammamatrices}) remain consistent with equations (97) and (98) of section (103) in \cite{Chandru} (comparing with our standard reference, \cite{Chandru}, we recover equation (99) in complex form). 

In order to represent spinorial objects (objects comprising spinors and gamma matrices) on a curved space-time, we use the following prescription on the tetrad formalism \cite{SVD_geometry_fields_cosmology}, viz. -- 
 Let $\mathcal{M}$ be a curved manifold with all conditions necessary for the existence of spin structure, and let U be a chart on $\mathcal{M}$  with coordinate functions ($x^{\alpha}$). Then, for representing spinorial objects, we (i) choose an orthonormal tetrad field $e^{\mu}_{(a)}(x^{\alpha})$ on $U$, (ii) define the Van der Waarden symbols $\sigma^{(a)}$ and $\tilde{\sigma}^{(a)}$ in this tetrad basis exactly as defined on Minkowski space in (\ref{vanderwaarden}) and choose a $\gamma$ representation (\ref{gammamatrices}); (iii) then, the $\sigma$'s in a local coordinate frame are then obtained via:
\begin{equation} \label{Van der Waarden symbols}
\sigma^{\mu}(x^{\alpha}) = e^{\mu}_{(a)}(x^{\alpha}) \sigma^{(a)} = \sqrt{2}\begin{bmatrix} l^{\mu} & m^{\mu} \\ \bar{m}^{\mu} & n^{\mu} \end{bmatrix} ~~~~~~~~~~~~~~~\tilde{\sigma}^{\mu} =e^{\mu}_{(a)}(x^\alpha)\tilde{\sigma}^{(a)} 
= \sqrt{2}\begin{bmatrix} n^{\mu} & -m^{\mu} \\ -\bar{m}^{\mu} & l^{\mu} \end{bmatrix} 
\end{equation}
with the $\gamma$ matrices obeying a similar transformation.

Thus, objects with world indices (containing world-indexed $\gamma$ matrices or spinors) are now functions of chosen orthonormal tetrads. These are defined \textit{a priori} in a local tetrad basis (with components identical to those defined on a flat Minkowski space-time) and \textit{then} carried into a curved space via the tetrads. This is unlike other geometrical world objects which are first defined naturally at a point in a manifold and subsequently carried to a local tangent space via tetrads.
We now aim to carry the Dirac equation (in NP) on $V_4$ into the $U_4$ space, building upon Section 102(d) of \cite{Chandru}. In order to calculate the covariant derivative of a spinor in $U_4$, we require the spinor affine connection coefficients. They are defined via the requirement that $\epsilon_{AB}$ and $\sigma$'s are covariantly constant. The analysis in \cite{Chandru} -- until Eq. 91 in the book -- still stands; however, the covariant derivatives are promoted to those acting on $U_4$. They are defined as follows::
\begin{align}
\nabla_{\mu}P^A &= \partial_{\mu} P^A + \Gamma^{A}_{\mu B} P^B  \label{eq:CD-Spinor}\\
\nabla_{\mu} \bar{Q}^{A'} &= \partial_{\mu}\bar{Q}^{A'} + \bar{\Gamma}^{A'}_{\mu B'} \bar{Q}^{B'} \label{eq:CD-CoSpinor}
\end{align}
\indent The $\Gamma$ terms here are added to the partial derivative when working with objects in $U_4$. Their values can completely be determined in terms of the spin coefficients, and we can readily evaluate its tetrad components using following formulae and the spin dyads \cite{SVD_geometry_fields_cosmology}):
\begin{equation}\label{eq:CDon spinors}
\Gamma^{A}_{\mu B} = \frac{1}{2} \sigma_{\nu}^{AY'}(\nabla_{\mu}\sigma^{\nu}_{BY'}) ~~~~~~~~~\bar{\Gamma}^{A'}_{\mu B'} = \frac{1}{2} \bar{\sigma}_{\nu}^{A'Y}(\bar{\nabla}_{\mu}\bar{\sigma}^{\nu}_{B'Y})
\end{equation} Using 
\textit{Friedman's lemma} (see pg. 542 of \cite{Chandru} for a full proof), we can express the various spin coefficients $\Gamma_{(a)(b)(c)(d')}$ in terms of covariant derivatives of the basis null vectors $l,n,m$ and $\bar{m}$. The covariant derivative here is exactly as defined in equation Eq. 3.3 (and explicitly written in Eq. 3.5) of \cite{jogia_Griffiths}. 

\indent Using this covariant derivative, it is readily seen how Eq. 95 and Eq. 96 in \cite{Chandru} get modified; viz. $\Gamma_{0000'} = \kappa^o + \kappa_1$ and $\Gamma_{1101'} = \mu^o + \mu_1$ (Naughts in the superscript are used to indicate the original spin coefficients defined on $V_4$). The 12 independent spin coefficients are calculated in terms of covariant derivatives of null vectors and defined in the following table\footnote{In the generic case, all 12 have contorsion spin coefficients} (\ref{tab:generic spin coefficients U4}):

\begin{equation}\label{tab:generic spin coefficients U4}
\Gamma_{(a)(b)(c)(d')} = \begin{tabular}{|l||*{3}{c|}}\hline
\diagbox{(c)(d')}{(a)(b)}
&\makebox[3em]{00}&\makebox[3em]{01 or 10}&\makebox[3em]{11}
\\\hline\hline
00' &$\kappa^o + \kappa_1 $&$\epsilon^o+ \epsilon_1$&$\pi^o + \pi_1$\\\hline
10' &$\rho^o + \rho_1$& $\alpha^o + \alpha_1$& $\lambda^o + \lambda_1$\\\hline
01' &$\sigma^o + \sigma_1$&$\beta^o+\beta_1$&$\mu^o +\mu_1$\\\hline
11' &$\tau^o+\tau_1$&$\gamma^o+\gamma_1$&$\nu^o + \nu_1$\\\hline\hline
\end{tabular}
\end{equation}

\subsection{Contorsion spin coefficients in terms of Dirac spinor components}\label{sec:Contorsion interms of spinor}

The spin density tensor of matter ($S^{\mu\nu\lambda}$) can be written as a world tensor in $U_4$ made up of the Dirac spinor, its adjoint, and gamma matrices: 
\begin{equation}\label{eq:Spin density of Dirac particles}
S^{\mu\nu\alpha} = \frac{-i\hbar c}{4}\bar{\psi}\gamma^{[\mu} \gamma^{\nu} \gamma^{\alpha]} \psi
\end{equation}
\indent The ECD field equations show that $T^{\mu\nu\alpha} = kS^{\mu\nu\alpha} $ where $T^{\mu\nu\alpha}$ is the modified torsion tensor defined in Eq. 2.3 of \cite{hehl_RMP}. It can be shown that, for Dirac field,  $T^{\mu\nu\alpha} =  -K^{\mu\nu\alpha} = kS^{\mu\nu\alpha}$ as in Eq. 5.6 of \cite{Hehl1971}. Here, $k$ is a gravitational coupling constant containing the length scale $l_1$, i.e., $k=\frac{8\pi l_1^2}{\hbar c}$. For the standard theory, $l_1=L_{Pl}$. Substituting (\ref{eq:Spin density of Dirac particles}) in the field equations, we obtain the following:
\begin{equation}\label{eq:calculating contorsion1}
K^{\mu\nu\alpha} = - kS^{\mu\nu\alpha} = 2i\pi l_1^2 \bar{\psi}\gamma^{[\mu} \gamma^{\nu} \gamma^{\alpha]} \psi
\end{equation} 
where the $\gamma^{\mu}$'s are those defined in (\ref{gammamatrices}), generalised with world indices using orthonormal tetrads. We subsequently rewrite $K^{\mu\nu\alpha}$ (of which only four independent components are excited by the Dirac field) in the NP formalism; i.e., in the null tetrad basis, as follows:
\begin{equation}\label{eq:calculating contorsion2}
K_{(i)(j)(k)} = e_{(i)\mu}e_{(j)\nu}e_{(k)\alpha} K ^{\mu\nu\alpha}
\end{equation} 
where $e_{(i)\mu} = (l_{\mu},n_{\mu},m_{\mu},\bar{m}_{\mu})$ for $i = 0,1,2,3$
To calculate the contorsion spin coefficients, we need to evaluate the contorsion tensor with world indices as defined in (\ref{eq:calculating contorsion1}). Consider the product $\gamma^{\alpha}\gamma^{\beta}\gamma^{\mu}$, which is defined as:
\begin{equation}
\gamma^{\alpha}\gamma^{\beta}\gamma^{\mu} =\begin{pmatrix} 0 & (\tilde{\sigma}^{\alpha})^*(\sigma^{\beta})^*(\tilde{\sigma}^{\mu})^* \\ (\sigma^{\alpha})^*(\tilde{\sigma}^{\beta})^*(\sigma^{\mu})^* & 0 \end{pmatrix}
\end{equation}
\indent The explicit form of this matrix is fairly expansive, and a full treatment is given in Appendix A. Eventually, we substitute in for the Dirac bispinor (as defined in \cite{Chandru}), and obtain the expressions for the contorsion spin coefficients in terms of the spinor components. We have, for example, for $\rho$ --
\begin{equation}
\rho = -K_{(0)(2)(3)} = -2\sqrt{2}i\pi l_1^2 [F_2\bar{F}_2 - G_1\bar{G}_1]
\end{equation}
\indent All the contorsion spin coefficients can be found in a similar fashion. After evaluating those, the eight non-zero spin coefficients excited by the Dirac spinor given in (\ref{diracspinor}) -- of which four are independent -- are as follows:
\begin{align} 
\tau_1 &= -2\beta_1 = K_{012} = 2\sqrt{2}i\pi  l_1^2 (F_2\bar{F}_1+ G_2\bar{G}_1) \label{eq:tau1} \\
\pi_1 &= -2\alpha_1 = K_{013} = 2\sqrt{2}i\pi  l_1^2 (-F_1\bar{F}_2 - G_1\bar{G}_2) \\
\mu_1 &= -2\gamma_1 = -K_{123} = 2\sqrt{2}i\pi  l_1^2 (F_1\bar{F}_1 - G_2\bar{G}_2) \\
\rho_1 &= -2\epsilon_1 = -K_{023} = 2\sqrt{2}i\pi  l_1^2 (G_1\bar{G}_1 - F_2\bar{F}_2) 
\label{eq:gamma1}
\end{align}
\indent From the above relations, we have:
\begin{align}
\mu_1 &= -\mu_1^*\\
\rho_1 &= -\rho_1^*\\
\pi_1 &= + \tau_1^*
\end{align}
\indent The table (\ref{tab:generic spin coefficients U4}) is modified as follows:
\begin{equation}\label{tab:Dirac spin coefficients U4}
\Gamma_{(a)(b)(c)(d')} = \begin{tabular}{|l||*{3}{c|}}\hline
\backslashbox{(c)(d')}{(a)(b)}
&\makebox[3em]{00}&\makebox[3em]{01 or 10}&\makebox[3em]{11}
\\\hline\hline
00' &$\kappa_0$ & $\epsilon_0 - \rho_1/2$&$\pi_0 + \pi_1$\\\hline
10' &$\rho_0 + \rho_1$& $\alpha_0 - \pi_1/2$& $\lambda_0$\\\hline
01' &$\sigma_0$&$\beta_0- \tau_1/2$&$\mu_0 +\mu_1$\\\hline
11' &$\tau_0+\tau_1$&$\gamma_0- \mu_1/2$&$\nu_0$\\\hline\hline
\end{tabular}
\end{equation}

Next, we formulate ECD theory in the NP formalism. There are three equations in this theory - the Dirac equation on $U_4$ (known as the Hehl-Datta equation), the gravitation field equation on $U_4$, and an algebraic equation relating torsion and spin. The algebraic equation is given in Eqn. (\ref{eq:calculating contorsion1}). In the next two sections, we formulate the Dirac equation and the gravitation field equations explicitly on $U_4$ respectively.

\subsection{The Dirac equation with torsion in the NP formalism} 

The Dirac equation on $U_4$ (also known as the Hehl-Datta equation) is: 
\begin{equation}
i\gamma^{\mu}\nabla_{\mu}\psi = \frac{mc}{\hbar}\psi = \frac{\psi}{2l_2}
\end{equation}
where $\nabla$ here denotes covariant derivative on $U_4$ and $l_2 = \frac{\lambda_c}{2}$ for standard theory. It can be written in the following matrix form:
\begin{align}
i\begin{pmatrix} 0 & (\tilde{\sigma}^{\mu})^* \\ (\sigma^{\mu})^* & 0 \end{pmatrix}\nabla_{\mu}\begin{pmatrix} P^A \\ \bar{Q}_{B'} \end{pmatrix} = \frac{1}{2\sqrt{2}l_2}\begin{pmatrix} P^A \\ \bar{Q}_{B'} \end{pmatrix}
\end{align}
\indent This can be written as a pair of matrix equations:
\begin{align}
\begin{pmatrix} \sigma^{\mu}_{00'}  & \sigma^{\mu}_{10'} \\ \sigma^{\mu}_{01'} & \sigma^{\mu}_{11'} \end{pmatrix}\nabla_{\mu}\begin{pmatrix} P^0 \\ P^1 \end{pmatrix} + \frac{i}{2\sqrt{2}l_2}\begin{pmatrix} -\bar{Q}^{1'}\\ \bar{Q}^{0'} \end{pmatrix} &= 0 \\
\begin{pmatrix} \sigma^{\mu}_{11'}  & -\sigma^{\mu}_{10'} \\ -\sigma^{\mu}_{01'} & \sigma^{\mu}_{00'} \end{pmatrix}\nabla_{\mu}\begin{pmatrix} -\bar{Q}^{1'}\\ \bar{Q}^{0'} \end{pmatrix} + \frac{i}{2\sqrt{2}l_2}\begin{pmatrix} P^0\\ P^1 \end{pmatrix} &= 0
\end{align}
\indent Working out explicitly, the first equation is:
\begin{equation}
\begin{split}
\frac{i}{2\sqrt{2}l_2}\bar{Q}^{1'} &= \sigma^{\mu}_{00'}\nabla_{\mu} P^0 + \sigma^{\mu}_{10'}\nabla_{\mu}P^1 = (\partial_{00'}P^0 + \Gamma^0_{~i00'}P^i) + (\partial_{10'}P^1 + \Gamma^1_{~i10'}P^i) \\
&=(D+\Gamma^0_{~000'}P^0 + \Gamma^0_{~100'}P^1) + (\delta^* + \Gamma^1_{~010'}P^0+ \Gamma^1_{~110'}P^1) \\
\Rightarrow \frac{i}{2\sqrt{2}l_2}G_1 &= (D+\epsilon_0 - \rho_0)F_1 + (\delta^{*} + \pi_0-\alpha_0)F_2 + \frac{3}{2}(\pi_1F_2 - \rho_1F_1)
\end{split}
\end{equation}
where we have used the gamma matrices as defined in (\ref{gammamatrices}), computed the covariant derivatives using (\ref{eq:CD-Spinor}), (\ref{eq:CD-CoSpinor}) and the spin connections in terms of contorsion spin coefficients as given in (\ref{tab:Dirac spin coefficients U4}). Using this procedure (a full treatment given in Appendix B), the four Dirac equations are obtained as:
\begin{align}
(D+\epsilon_0 - \rho_0)F_1 + (\delta^{*} + \pi_0-\alpha_0)F_2 + \frac{3}{2}(\pi_1F_2 - \rho_1F_1) &= i b(l_2)G_1 \label{eq:Dirac eq U4 NP1.1}\\
(\Delta + \mu_0 - \gamma_0)F_2 + (\delta +\beta_0 -\tau_0)F_1 +
\frac{3}{2}(\mu_1 F_2 - \tau_1 F_1) &= i b(l_2)G_2 \\
(D+\epsilon_0^* - \rho_0^*)G_2 - (\delta + \pi_0^*-\alpha_0^*)G_1 - \frac{3}{2}(\tau_1G_1 - \rho_1 G_2 ) &= i b(l_2)F_2\\
(\Delta + \mu_0^* - \gamma_0^*)G_1 - (\delta^* +\beta_0^* -\tau_0^*)G_2 -
\frac{3}{2}(\mu_1 G_1 - \pi_1 G_2) &=i b(l_2)F_1 \label{eq:Dirac eq U4 NP1.4}
\end{align}
\indent Substituting in the spinorial form of the contorsion spin coefficients in (\ref{eq:tau1}) - (\ref{eq:gamma1}), we obtain: 
\begin{align}
(D+\epsilon_0 - \rho_0)F_1 + (\delta^{*} + \pi_0-\alpha_0)F_2 + ia(l_1)[(-F_1\bar{F}_2 - G_1\bar{G}_2)F_2 + (F_2\bar{F}_2 - G_1\bar{G}_1)F_1] &= i b(l_2)G_1 \label{eq:Dirac eq U4 NP2.1}\\
(\Delta + \mu_0 - \gamma_0)F_2 + (\delta +\beta_0 -\tau_0)F_1+ ia(l_1)[(F_1\bar{F}_1 - G_2\bar{G}_2) F_2 -(F_2\bar{F}_1+ G_2\bar{G}_1)F_1] &= i b(l_2)G_2 \\
(D+\epsilon_0^* - \rho_0^*)G_2 - (\delta + \pi_0^*-\alpha_0^*)G_1 -  ia(l_1)[(F_2\bar{F}_2 - G_1\bar{G}_1) G_2 + (F_2\bar{F}_1+ G_2\bar{G}_1)G_1]  &= i b(l_2)F_2\\
(\Delta + \mu_0^* - \gamma_0^*)G_1 - (\delta^* +\beta_0^* -\tau_0^*)G_2 - ia(l_1)[(F_1\bar{F}_1 - G_2\bar{G}_2) G_1 -(-F_1\bar{F}_2 - G_1\bar{G}_2) G_2]&= i b(l_2) F_1 \label{eq:Dirac eq U4 NP2.4}
\end{align}
where $a(l_1) = 3\sqrt{2}\pi l_1^2$ and $b(l_2) = \frac{1}{2\sqrt{2}l_2}$.\\ 
\indent These equations can be condensed into the following form:
\begin{align}
(D+\epsilon_0 - \rho_0)F_1 + (\delta^{*} + \pi_0-\alpha_0)F_2 &= i[b(l_2)+a(l_1)\xi]G_1\label{1-DEonU_4final}\\
(\Delta + \mu_0 - \gamma_0)F_2 + (\delta +\beta_0 -\tau_0)F_1 &= i[b(l_2)+a(l_1)\xi]G_2 \\
(D+\epsilon_0^* - \rho_0^*)G_2 - (\delta + \pi_0^*-\alpha_0^*)G_1 &= i[b(l_2)+a(l_1)\xi^*]F_2\\
(\Delta + \mu_0^* - \gamma_0^*)G_1 - (\delta^* +\beta_0^* -\tau_0^*)G_2  &= i[b(l_2)+a(l_1)\xi^*]F_1\label{4-DEonU_4final}
\end{align}
where $\xi = F_1\bar{G}_1 + F_2\bar{G}_2$ and $
\xi^* = \bar{F}_1 G_1 + \bar{F}_2 G_2$. These equations should be compared and contrasted with the torsionless Dirac equations in \cite{Chandru}, and then we see that the impact of torsion is to include the term $a\xi$ on the right hand side of the first two equations, and $a\xi^*$ in the last two equations.

\subsection{\texorpdfstring{The gravitation equations on $U_4$ in NP formalsim}
	{The dynamical EM tensor Tmunu and spin density in the NP formalism}}\label{subsec:Tmunu}
The equation of interest here is (\ref{eq:efeu4}), reproduced here: 
\begin{equation}
G_{\mu\nu}(\{\}) = \frac{8\pi l_1^2}{\hbar c} T_{\mu\nu} - \frac{1}{2} \bigg{(}\frac{8\pi l_1^2}{\hbar c}\bigg{)}^2 g_{\mu\nu}  S^{\alpha\beta\lambda}S_{\alpha\beta\lambda}
\end{equation}
\indent On the left hand side, we have $G_{\mu\nu}(\{\})$, which has been completely evaluated in the NP formalism in \cite{Chandru}. There are two terms on right hand side -- the first of these is the metric energy-momentum tensor ($T_{\mu\nu}$) formulated on $U_4$ and is given by equation (\ref{dynamic EM tensor}). In what follows, we will give a prescription to compute the various components of $T_{\mu\nu}$, under the definition:
\begin{equation}\label{eq1}
T_{\mu\nu} = \frac{i\hbar c}{4}\Big[\bar{\psi}\gamma_{\mu} \nabla^{\{\}}_{\nu}\psi + \bar{\psi}\gamma_{\nu} \nabla^{\{\}}_{\mu}\psi - \nabla^{\{\}}_{\mu}\bar{\psi} \gamma_{\nu}\psi -\nabla^{\{\}}_{\nu}\bar{\psi} \gamma_{\mu}\psi  \Big]
\end{equation}
\indent First, we choose a tetrad basis and construct Van der Waarden symbols as defined in (\ref{Van der Waarden symbols}). Using these, we construct Dirac gamma matrices in the complex Weyl representation as defined in (\ref{gammamatrices}). Now, the expression for the covariant derivatives of spinors -- see (\ref{eq:CD-Spinor}),(\ref{eq:CD-CoSpinor}),(\ref{eq:CDon spinors}) -- can be expressed in terms of the gamma matrices, yielding:
\begin{equation}
\begin{split}
T_{\mu\nu} = \frac{i\hbar c}{4}\Big[&\bar{\psi}\gamma_{\mu} \partial_{\nu}\psi + \frac{1}{4}\bar{\psi}(\gamma_{\mu}\gamma^{\alpha}\nabla^{\{\}}_{\nu}\gamma_{\alpha})\psi + \bar{\psi}\gamma_{\nu} \partial_{\mu}\psi + \frac{1}{4}\bar{\psi}(\gamma_{\nu}\gamma^{\alpha}\nabla^{\{\}}_{\mu}\gamma_{\alpha})\psi \\
&-\partial_{\mu}\bar{\psi} \gamma_{\nu}\psi - \frac{1}{4}(\bar{\gamma}^{\alpha}\bar{\nabla}^{\{\}}_{\mu}\bar{\gamma}_{\alpha})\bar{\psi}\gamma_{\nu}\psi-\partial_{\nu}\bar{\psi} \gamma_{\mu}\psi  - \frac{1}{4}(\bar{\gamma}^{\alpha}\bar{\nabla}^{\{\}}_{\nu}\bar{\gamma}_{\alpha})\bar{\psi}\gamma_{\mu}\psi\Big]
\end{split}
\end{equation}
\indent Here, the gamma matrices and other variables are expressed in the basis of null vectors $l,n,m$ and $\bar{m}$. For the generic metric energy-momentum tensor $T_{\mu\nu}$, no further simplification is possible. The expression for $T_{\mu\nu}$ in the NP formalism will however simplify under certain symmetries or specific conditions that the system in question is subjected to. For example, if the background metric is $\eta_{\mu\nu}$, then (for illustration purposes) the $T_{12}$ component of metric EM tensor is given by:
\begin{equation}
\begin{split}
T_{12}^{\text{(NP)}} = \frac{i\hbar c}{4 \sqrt{2}} \bigg{(}& i\bar{F}_2 (\delta+\delta^*)F_1 - i \bar{F}_1(\delta+\delta^*) F_2 - i \bar{G}_2(\delta+\delta^*) G_1 + i \bar{G}_1(\delta+\delta^*) G_2 \\& -i\bar{F}_2 (\delta-\delta^*)F_1 -i \bar{F}_1(\delta-\delta^*) F_2 + i\bar{G}_2 (\delta-\delta^*)G_1 + i\bar{G}_1(\delta-\delta^*) G_2 \bigg{)} \\ & -i(\delta+\delta^*)\bar{F}_2 F_1 + (\delta+\delta^*) i \bar{F}_1 F_2 + (\delta+\delta^*) i \bar{G}_2 G_1 -(\delta+\delta^*) i \bar{G}_1 G_2 \\& + (\delta-\delta^*)i\bar{F}_2F_1 + (\delta-\delta^*)i \bar{F}_1 F_2 -(\delta-\delta^*) i\bar{G}_2 G_1 -(\delta-\delta^*) i\bar{G}_1 G_2 \bigg{)}  
\end{split}
\end{equation}
\indent With this prescription, we are able to evaluate all the components of $T_{\mu\nu}$, achieving a particularly simple form in the case of a Minkowskian background metric.

In (\ref{eq:efeu4}), we also have an additional term in terms of the spin density tensor, given as $\frac{4\pi l_1^2}{\hbar c}g_{\mu\nu} S^{\alpha\beta\lambda}S_{\alpha\beta\lambda}$. Using our expression for the spin density, we can evaluate this term: 
\begin{align}
	\frac{4\pi l_1^2}{\hbar c}g_{\mu\nu} S^{\alpha\beta\lambda}S_{\alpha\beta\lambda} &= \frac{-\pi l_1^2\hbar c}{4}\Big(\bar{\psi}\gamma^{[\alpha} \gamma^{\beta} \gamma^{\lambda]} \psi  \Big) \Big(\bar{\psi}\gamma_{[\alpha} \gamma_{\beta} \gamma_{\lambda]} \psi  \Big) \\
	&=  \frac{-\pi l_1^2\hbar c}{4}\Big(\bar{\psi}\gamma^{[(i)} \gamma^{(j)} \gamma^{(k)]} \psi  \Big) \Big(\bar{\psi}\gamma_{[(i)} \gamma_{(j)} \gamma_{(k)]} \psi  \Big)) \\
	&=6\pi \hbar c l_1^2g_{\mu\nu} (F_1\bar{G}_1 + F_2\bar{G}_2)(\bar{F}_1 G_1 + \bar{F}_2 G_2)\\
	&= 6\pi \hbar c l_1^2g_{\mu\nu} \xi\xi^*\\
	&= 12\pi \hbar c l_1^2 (  l_{(\mu}n_{\nu)} - m_{(\mu}\bar{m}_{\nu)}) \xi\xi^*
\end{align}
i.e., we find that it turns out to be proportional to the $\xi$ parameter introduced.

This completes the formulation of the Einstein-Cartan-Dirac equations in the NP formalism. The formalism can be used to
examine how torsion modifies the properties of the Einstein-Dirac system.
Next, we investigate some solutions of the Hehl-Datta equations. In future work we hope to extend these studies to Poincar\'e gauge gravity with propagating torsion.

\section{Solutions to HD equations in Minkowski space}
\subsection{Motivation}

In the previous section, we formulated the ECD equations in the NP formalism. In this section, we aim to solve them. The simplest space-time with torsion is the Minkowski ($\eta_{\mu\nu}$) space-time with a manifold that has non-zero torsion. In this space-time, the Dirac equation on $U_4$ looks very similar to the linear Dirac equation with modified mass (the torsion-related term which modifies it is bilinear in the Dirac states). In this spirit, we will consider modifications (due to torsion) to well-studied solutions to the linear Dirac equation (e.g. plane wave solutions).

In addition, there are good (physical) reasons to work within Minkowski space-time, to find solution(s) of the HD equations incorporating torsion. In a recent work \cite{TP_1,TP_2,GRFessay2018}, a duality between large and small masses (correspondingly, between Riemannian curvature and torsion) has been proposed, explicitly constructed in the ``curvature-torsion duality conjecture" therein. For this conjecture to hold true, a solution to Dirac equation on Minkowski space with torsion must exist -- along with certain other conditions. One such additional condition is the vanishing of the $(T-S)_{\mu\nu}$ tensor, as defined in Appendix C.

While we proceed in the following section to find solutions to the HD equations on Minkowski space for their own sake, the reader may find, in \cite{GRFessay2018}, useful extensions to this work. To this end, in the Appendices (ref. Appendix C) we have also computed the $(T-S)_{\mu\nu}$ tensor in certain cases, for completeness.

\subsection{The Hehl-Datta equations on Minkowski space with torsion}

The HD equations on Minkowski space with torsion (in the NP formalism) are as follows:

\begin{align}
D F_1 + \delta^{*}F_2 &= i[b(l_2)+a(l_1)\xi]G_1 \label{eq:HD_in_NP_minkowski1}\\
\Delta F_2 + \delta F_1 &= i[b(l_2)+a(l_1)\xi]G_2  \label{eq:HD_in_NP_minkowski2}\\
DG_2 - \delta G_1 &= i[b(l_2)+a(l_1)\xi^*]F_2 \label{eq:HD_in_NP_minkowski3}\\
\Delta G_1 - \delta^* G_2 &= i[b(l_2)+a(l_1)\xi^*]F_1 \label{eq:HD_in_NP_minkowski4}
\end{align}

In a Cartesian coordinate system $(ct,x,y,z)$\footnote{Setting $c=1$ by convention} we have:
\begin{align}
(\partial_0+\partial_3)F_1 + (\partial_1+i\partial_2) F_2 = i\sqrt{2}[b(l_2)+a(l_1)\xi]G_1 \label{eq:HD_in_cartesian_minkowski1}\\
(\partial_0-\partial_3)F_2 + (\partial_1-i\partial_2) F_1 = i\sqrt{2}[b(l_2)+a(l_1)\xi]G_2\label{eq:HD_in_cartesian_minkowski2}\\
(\partial_0+\partial_3)G_2 - (\partial_1-i\partial_2) G_1 = i\sqrt{2}[b(l_2)+a(l_1)\xi^*]F_2 \label{eq:HD_in_cartesian_minkowski3}\\
(\partial_0-\partial_3)G_1 - (\partial_1+i\partial_2) G_2 = i\sqrt{2}[b(l_2)+a(l_1)\xi^*]F_1 \label{eq:HD_in_cartesian_minkowski4}
\end{align}

In cylindrical polar coordinates $(ct,r,\phi,z)$, we have:

\begin{align}
r\partial_t F_1 + e^{i\phi} r\partial_r F_2  + ie^{i\phi}\partial_{\phi} F_2 + r\partial_z F_1 &= ir\sqrt{2} [b(l_2)+a(l_1)\xi]G_1 \label{eq:HD_in_cylindrical_minkowski1} \\
r\partial_t F_2 + e^{-i\phi} r\partial_r F_1  - ie^{-i\phi}\partial_{\phi} F_1 - r\partial_z F_2&= ir\sqrt{2} [b(l_2)+a(l_1)\xi]G_2 \label{eq:HD_in_cylindrical_minkowski2}\\
r\partial_t G_2 - e^{-i\phi}r\partial_r G_1  + ie^{-i\phi}\partial_{\phi} G_1 + cr\partial_z G_2&= ir\sqrt{2} [b(l_2)+a(l_1)\xi^*]F_2 \label{eq:HD_in_cylindrical_minkowski3}\\
r\partial_t G_1 - e^{i\phi} r\partial_r G_2 - ie^{i\phi} \partial_{\phi} G_2 - r\partial_z G_1&= ir\sqrt{2} [b(l_2)+a(l_1)\xi^*]F_1 \label{eq:HD_in_cylindrical_minkowski4}
\end{align}

Likewise, in spherical polar coordinates $(ct,r,\theta, \phi)$: 
\small
\begin{align}
\partial_t F_1+\cos{\theta} \partial_r F_1 - \frac{\sin{\theta}}{r}\partial_{\theta}F_1 + \frac{ie^{i\phi}}{r\sin{\theta}}\partial_{\phi}F_2 + e^{i\phi}\sin{\theta} \partial_rF_2  +\frac{ e^{i\phi}\cos{\theta}}{r}\partial_{\theta}F_2&= i\sqrt{2}[b(l_2)+a(l_1)\xi]G_1 \label{eq:HD_in_SPC_minkowski1}\\
\partial_t F_2-\cos{\theta} \partial_r F_2 - \frac{\sin{\theta}}{r}\partial_{\theta}F_2 + \frac{ie^{-i\phi}}{r\sin{\theta}}\partial_{\phi}F_1 + e^{-i\phi}\sin{\theta} \partial_rF_1 -\frac{ e^{-i\phi}\cos{\theta}}{r}\partial_{\theta}F_1 &= i\sqrt{2}[b(l_2)+a(l_1)\xi]G_2\\
\partial_t G_2+\cos{\theta} \partial_r G_2 - \frac{\sin{\theta}}{r}\partial_{\theta}G_2  - \frac{ie^{-i\phi}}{r\sin{\theta}}\partial_{\phi}G_1 - e^{-i\phi}\sin{\theta} \partial_rG_1 +\frac{ e^{-i\phi}\cos{\theta}}{r}\partial_{\theta}G_1  &= i\sqrt{2}[b(l_2)+a(l_1)\xi^*]F_2 \\
\partial_t G_1-\cos{\theta} \partial_r G_1 - \frac{\sin{\theta}}{r}\partial_{\theta}G_1 -\frac{ie^{i\phi}}{r\sin{\theta}}\partial_{\phi}G_2 - e^{i\phi}\sin{\theta} \partial_rG_2  -\frac{ e^{i\phi}\cos{\theta}}{r}\partial_{\theta}G_2&= i\sqrt{2}[b(l_2)+a(l_1)\xi^*]F_1 \label{eq:HD_in_SPC_minkowski4}
\end{align}
\normalsize
\subsection{A  non-static solution  in 1+1 dimensions}
In the following analysis, we will assume an ansatz of the form $F_1 = G_2 $ and $F_2 = G_1 $, and further assume that the Dirac states are a function of only $t$ and $z$. The four equations -- in Cartesian (\ref{eq:HD_in_cartesian_minkowski1}) - (\ref{eq:HD_in_cartesian_minkowski4}) as well as cylindrical polar coordinates (\ref{eq:HD_in_cylindrical_minkowski1}) - (\ref{eq:HD_in_cylindrical_minkowski4})) -- reduce to the following two independent equations\footnote{We note that $\xi = 2 Re(F_1\bar{F}_2)$, thus $\xi = \xi^*$. Furthermore, $a$ and $b$ are henceforth shorthand for $a(l)$ and $b(l)$.} 
\begin{equation}
\begin{split}
	\partial_t \psi_1 + \partial_z \psi_2 - i\sqrt{2} b \psi_1+\frac{ia}{\sqrt{2}}(|\psi_2|^2-|\psi_1|^2)\psi_1 &= 0\\
	\partial_t \psi_2 + \partial_z \psi_1 +  i\sqrt{2}b\psi_2 +\frac{ia}{\sqrt{2}}(|\psi_1|^2-|\psi_2|^2)\psi_2 &= 0
\end{split}
\end{equation}
where $\psi_1 = F_1 + F_2$ and $\psi_2 = F_1 - F_2$. If we were to define  $\sqrt{2}b \equiv -m$ and $a = 2\sqrt{2}\lambda$, we will get:
\begin{equation}
\begin{split}
	\partial_t \psi_1 + \partial_z \psi_2 +im\psi_1 +2i\lambda(|\psi_2|^2-|\psi_1|^2)\psi_1 &= 0\\
	\partial_t \psi_2 + \partial_z \psi_1 -im\psi_2 +2i\lambda(|\psi_1|^2-|\psi_2|^2)\psi_2 &= 0
\end{split}
\end{equation}
\indent These equations are identical to those studied in \cite{alvarez1983numerical}, which investigates the convergence and stability of the difference scheme for the non-linear Dirac equation in $1+1$ dimensions. Proceeding as in \cite{alvarez1983numerical}, we use the following solitary wave ansatz:
\begin{equation}
	\psi = \begin{pmatrix} \psi_1\\ \psi_2 \end{pmatrix} = \begin{pmatrix} A(z)\\ iB(z)\end{pmatrix} e^{-i\Lambda t}
\end{equation}
where $A(z)$ and $B(z)$ are real functions. Substituting in, we have:
\begin{equation}
\begin{split}
	B' - (\sqrt{2}b + \Lambda)A - \frac{a}{\sqrt{2}}(A^2-B^2)A &= 0 \\
	A' - (\sqrt{2}b - \Lambda)B - \frac{a}{\sqrt{2}}(A^2-B^2)B &= 0
\end{split}
\end{equation}
which admits the following solutions:
\begin{align} 
	A(z) &= \frac{-i2^{3/4} (\sqrt{2} b-\Lambda)}{\sqrt{a}} \frac{\sqrt{(\sqrt{2}b+\Lambda)} \cosh(z \sqrt{2 b^2-\Lambda ^2})}{[\Lambda  \cosh (2z\sqrt{2 b^2-\Lambda ^2}) - \sqrt{2} b]} \label {eq:1+1_A(z)}\\
	B(z) &= \frac{-i2^{3/4} (\sqrt{2} b+\Lambda)}{\sqrt{a}}\frac{\sqrt{(\sqrt{2}b-\Lambda)} \sinh(z \sqrt{2 b^2-\Lambda ^2})}{[\Lambda  \cosh (2z\sqrt{2 b^2-\Lambda ^2}) - \sqrt{2} b]} \label {eq:1+1_B(z)}
\end{align}
\indent It can be seen upon the substitutions $\lambda = 0.5$ (equivalently $a = \sqrt{2}$) and $m = 1$ (equivalently $m_0 = -1$), that this is a generalisation of the equations for $A(z)$ and $B(z)$ in \cite{alvarez1983numerical}(see section III). A similar solution is found in \cite{zecca}, with $a(l_1) = a(L_{pl})$ and $b(l_2) = b(\lambda_c)$. In terms of the spinor components:
\small
\begin{align}
F_1&= G_2 =  \frac{\sqrt{(2b^2-\Lambda^2)}}{2}\bigg{[}\frac{-i2^{3/4}}{\sqrt{a}} \frac{\sqrt{(\sqrt{2}b-\Lambda)} \cosh(z \sqrt{2 b^2-\Lambda ^2})}{[\Lambda  \cosh (2z\sqrt{2 b^2-\Lambda ^2}) - \sqrt{2} b]} + \frac{2^{3/4}}{\sqrt{a}}\frac{\sqrt{(\sqrt{2}b+\Lambda)} \sinh(z \sqrt{2 b^2-\Lambda ^2})}{[\Lambda  \cosh (2z\sqrt{2 b^2-\Lambda ^2}) - \sqrt{2} b]} \bigg{]} e^{-i\Lambda t} \\
F_2&= G_1 =  \frac{\sqrt{(2b^2-\Lambda^2)}}{2}\bigg{[}\frac{-i2^{3/4} }{\sqrt{a}} \frac{\sqrt{(\sqrt{2}b-\Lambda)} \cosh(z \sqrt{2 b^2-\Lambda ^2})}{[\Lambda  \cosh (2z\sqrt{2 b^2-\Lambda ^2}) - \sqrt{2} b]} - \frac{2^{3/4} }{\sqrt{a}}\frac{\sqrt{(\sqrt{2}b+\Lambda)} \sinh(z \sqrt{2 b^2-\Lambda ^2})}{[\Lambda  \cosh (2z\sqrt{2 b^2-\Lambda ^2}) - \sqrt{2} b]} \bigg{]} e^{-i\Lambda t} 
\end{align}
\normalsize
and the parameter $\xi$ characterising torsion takes the form:
\begin{equation}
\xi =  \dfrac{-2\sqrt{2}(2b^2-\Lambda^2)(\sqrt{2}b-\Lambda\cosh (2z\sqrt{2 b^2-\Lambda ^2} )}{a[\Lambda  \cosh (2z\sqrt{2 b^2-\Lambda ^2}) - \sqrt{2}b]^2}
\end{equation}
\indent The probability density is given by the zeroth component of the four-vector fermion current $j^0 = \bar{\psi}\gamma^0\psi = \psi^{\dagger} \psi = 2\Big( |F_1|^2+|F_2|^2 \Big) = \Big( |A|^2+|B|^2 \Big)$
For the subsequent analysis, we define the following dimensionless variables:
\begin{equation}
\begin{split}
p &= \sqrt{2}bz \\
w &=  - \frac{\Lambda}{\sqrt {2} b} \\
\tilde{A}(p) &= \frac{\sqrt{a}}{2\sqrt{b}} A(z) \\
\tilde{B}(p) &= \frac{\sqrt{a}}{2\sqrt{b}} B(z)\\
\tilde{j}^0 &= \frac{a}{4b} j^0  = 0
\end{split}
\end{equation}
\indent With these definitions, we have $[p]=[w]=[\tilde{A}(p)]=[\tilde{B}(p)]=[\tilde{j}^0]=0$; i.e., all these quantities are now dimensionless. Scaled thus, $A(p)$ and $B(p)$ take the form:
\begin{align} 
A(p) &= \frac{2i(1+w)}{\sqrt{a}} \frac{\sqrt{b(1-w)} \cosh(p\sqrt{1-w^2})}{(w\cosh (2p\sqrt{1-w^2}) +1)} \label {eq:1+1_A(p)}\\
B(p) &= \frac{2i(1-w)}{\sqrt{a}}\frac{\sqrt{b(1+w)} \sinh(p\sqrt{1-w^2})}{(w\cosh (2p\sqrt{1-w^2}) +1)} \label {eq:1+1_B(p)}
\end{align}
\indent There are six unique cases (corresponding to values of $w$) which give different solutions. In each case, we will consider torsion-less limit (the linear Dirac equation) in order to compare and contrast the behaviour. The equations and plots for the linear case can be found in Appendix D.\\ 

\noindent\underline{Case I}: $w \in (-\infty,-1)$: 
The equations reduce to:
\begin{gather} 
\tilde{A}(p) = i(1+w) \frac{\sqrt{(|w|+1)} \cos(p\sqrt{w^2-1})}{(1- |w|\cos (2p\sqrt{w^2-1}))} \label {eq:1+1_A(p)case1}\\
\tilde{B}(p) = i(w-1)\frac{\sqrt{(|w|-1)} \sin(p\sqrt{w^2-1})}{(1- |w|\cos (2p\sqrt{w^2-1})) } \label {eq:1+1_B(p)case1}\\
\tilde{j}^0 = \Bigg[\frac{(w+1)^2(|w|+1)\cos^2(p\sqrt{w^2-1}) +(w-1)^2(|w|-1)\sin^2(p\sqrt{w^2-1})}{(1- |w|\cos (2p\sqrt{w^2-1}))^2}  \Bigg]\\
\end{gather}
\textit{Comments}: This solution has an infinite number of singularities placed periodically at non-zero values of $p$, and is clearly unphysical. An example of this case (with $w=-2$) can be seen in the left column of Fig. \ref{fig_case1_6}.

\textit{Comparison with torsionless case}: For $w \in (-\infty,-1)$, the linear Dirac equation gives plane waves solutions, which are physically meaningful, and the probability density fluctuates sinusoidally. It is the addition of torsion that makes this case unphysical. A plot has been made (for $w=-2$) in Fig. \ref{fig:nontorsion Dirac case}.\\

\noindent\underline{Case II}: $w = \pm 1$ (trivial case):
The equations reduce to:
\begin{equation} 
\tilde{A}(p) = 0 ~~~~~~~
\tilde{B}(p) = 0 ~~~~~~~
\tilde{j}^0 = 0
\end{equation}

\noindent\underline{Case III}: $w \in (-1,0)$:
The equations reduce to:
\begin{gather} 
\tilde{A}(p) = i(1+w) \frac{\sqrt{(1+|w|)} \cosh(p\sqrt{1-w^2})}{(1-|w|\cosh (2p\sqrt{1-w^2}))} \label {eq:1+1_A(p)case3}\\
\tilde{B}(p) = i(1-w)\frac{\sqrt{(1-|w|)} \sinh(p\sqrt{1-w^2})}{(1-|w|\cosh (2p\sqrt{1-w^2}))} \label {eq:1+1_B(p)case3} \\
\tilde{j}^0 =\Bigg[\frac{(w+1)^2(|w|+1)\cosh^2(p\sqrt{1-w^2}) +(1-w)^2(1-|w|)\sinh^2(p\sqrt{1-w^2})}{(1- |w|\cosh (2p\sqrt{1-w^2}))^2}  \Bigg]
\end{gather}
\textit{Comments}: This solution has two singularities placed symmetrically around the origin at two finite (non-zero) values of $p$. In the infinite limit, it decays to zero. However, owing to the presence of singularities, we may still conider it an unphysical solution. An example (with $w=-0.5$) can be seen in the left column of Figure. \ref{fig_case3_4}\\
\textit{Comparison with torsionless case}: For $w \in (-1,0)$ the linear Dirac equation has unphysical solutions. The solutions grow exponentially to infinity as $p \rightarrow \pm\infty$. For $w=-0.5$, this solution is plotted in Fig. \ref{fig:nontorsion Dirac case}. As can be seen, for this case, both the linear (torsionless) and non-linear (with torsion) Dirac equations give unphysical solutions.\\

\noindent\underline{Case IV}: $w =0$:
The equations reduce to:
\begin{gather} 
\tilde{A}(p) = i \cosh(p) \\
\tilde{B}(p) = i \sinh(p) \\
\tilde{j}^0 =  [\cosh^2(p) + \sinh^2(p)]\label {eq:1+1_A(p)case4}
\end{gather}
\textit{Comments}: This solution blows up exponentially as $p \rightarrow \pm\infty$. Thus, it is clearly unphysical. This case (with $w=0$ has been plotted in the right column of Fig. \ref{fig_case3_4}

\textit{Comparison with torsionless case}: For $w =0$, the linear Dirac equation is unphysical. The solutions exponentially increase to infinity as $p \rightarrow +\infty$. A plot of the solutions (for $w=0$) is available in Fig. \ref{fig:nontorsion Dirac case}. Thus, for this case, both the linear and non-linear Dirac equations give unphysical solutions.\\

\noindent\underline{Case V}: $w \in (0,1)$:
The equations reduce to:
\begin{gather} 
\tilde{A}(p) = i(1+w) \frac{\sqrt{(1-w)} \cosh(p\sqrt{1-w^2})}{(1+w\cosh (2p\sqrt{1-w^2}))} \label {eq:1+1_A(p)case5}\\
\tilde{B}(p) = i(1-w)\frac{\sqrt{(1+w)} \sinh(p\sqrt{1-w^2})}{(1+ w\cosh (2p\sqrt{1-w^2}))} \label {eq:1+1_B(p)case5} \\
\tilde{j}^0 =  \Bigg[\frac{(1+w)^2(1-w)\cosh^2(p\sqrt{1-w^2}) +(1-w)^2(1+w)\sinh^2(p\sqrt{1-w^2})}{(1+ w\cosh (2p\sqrt{1-w^2}))^2}  \Bigg]
\end{gather}
\textit{Comments}: In this case, we have no singularities anywhere. All the functions (including the probability density) asymptotically vanish. Therefore, this case represents a physically viable solution. Depending on the exact nature of solution, we can consider two sub-cases: (a) with $w \in (0,\frac{1}{3})$ and (b) with $w \in [\frac{1}{3},1)$.\\
We see that (a) has a local minimum at the origin and two global maxima symmetric around the origin at non-zero $p$. A plot is provided in Fig. \ref{fig_case5} (in blue). On the other hand, (b) has global maximum at the origin and monotonically decays to zero at infinity. Two examples of this can be seen in Fig. \ref{fig_case5} (in orange and green). The solution for case (b) resembles a `blob'; further analysis of this can be found in the discussion. 

\textit{Comparison with torsionless case}: For $w \in (0,1)$ the linear Dirac equation gives unphysical solutions. The solutions increase exponentially to infinity as $p \rightarrow \pm\infty$. A plot of this solution (with $w=0.5$) can be seen in Fig. \ref{fig:nontorsion Dirac case}. The addition of torsion, as seen, makes the solutions physically meaningful.\\

\noindent\underline{Case VI}: $w \in (1, \infty)$:
The equations reduce to:
\begin{gather} 
\tilde{A}(p) = -(1+w) \frac{\sqrt{(w-1)} \cos(p\sqrt{w^2-1})}{(1+ w\cos (2p\sqrt{w^2-1}) } \label {eq:1+1_A(p)case6}\\
\tilde{B}(p) = -(1-w)\frac{\sqrt{(w+1)}\sin(p\sqrt{w^2-1})}{(1+ w\cos (2p\sqrt{w^2-1}) } \label {eq:1+1_B(p)case6}\\
\tilde{j}^0 =  \Bigg[\frac{(1+w)^2(w-1)\cos^2(p\sqrt{w^2-1}) +(1-w)^2(w+1)+\sin^2(p\sqrt{w^2-1})}{(1+ w\cos (2p\sqrt{w^2-1}))^2}  \Bigg]
\end{gather}
\textit{Comments}: This solution has an infinite number of singularities placed periodically over non-zero values of $p$, and is thus clearly unphysical. A plot (with $w=2$) is given in the left column of Fig. \ref{fig_case1_6}

\textit{Comparison with torsionless case}: For $w \in (1, \infty)$ the linear Dirac equation gives (physically meaningful) plane waves solutions. The probability density fluctuates sinusoidally. The addition of torsion makes this solution ultimately unphysical. A plot (with $w=2$) is available in Fig. \ref{fig:nontorsion Dirac case}.

\newpage

\noindent The following table summarises the various cases:
\bigskip
\begin{center}
\begin{tabular}{|| c || p{5cm} | p{5cm} ||}\hline
	\textbf{Cases} & \textbf{Solution(s) of the linear Dirac equation} & \textbf{Solution(s) of the Dirac equation with torsion} \\ \hline\hline
	Case I & Physical (Plane wave)  & Unphysical (infinite singularities)  \\
	\hline
	Case II & Trivial solution &Trivial solution   \\ \hline
	Case III & Unphysical (blows up exponentially at infinity) & Unphysical, (two singularities)   \\ \hline
	Case IV & Unphysical (blows up exponentially at infinity) & Unphysical (blows up exponentially at infinity)   \\ \hline
	Case V & Unphysical (blows up exponentially at infinity) & Physical (No singularity)   \\ \hline
	Case VI & Physical (Plane wave)  & Unphysical (infinite singularities)  \\ \hline\hline
\end{tabular}
\end{center}

\begin{figure}[h!] 
	\begin{center}
		\includegraphics[width=15cm, height=15cm]{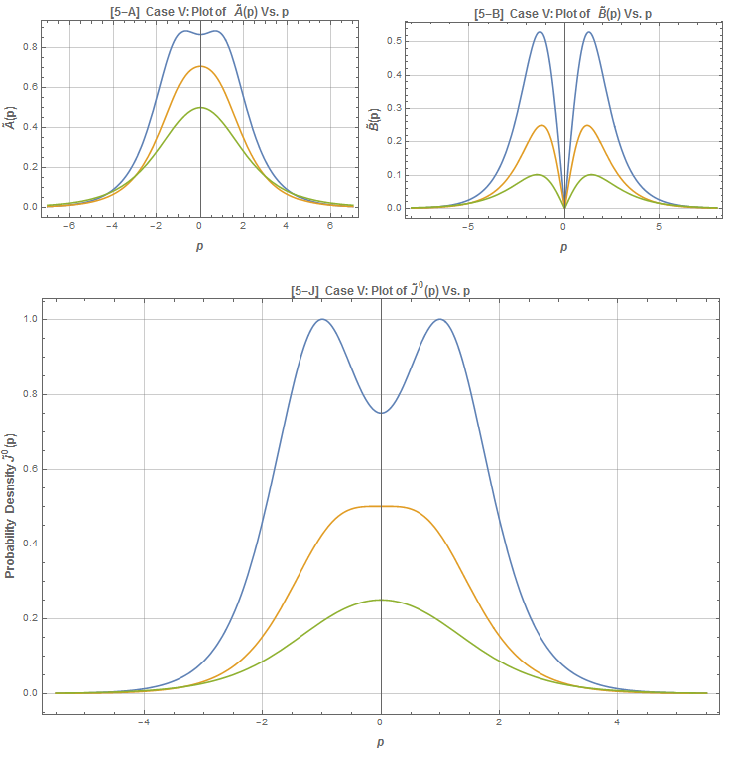}
		\caption{\textbf{Case V}.
			In all plots: \textit{Green}: Case V(a) with w=0.75], \textit{Orange}: Case V(a) with $w=0.5$,  \textit{Blue}: Case V(b) with $w=0.25$. Case V(a) has global maxima at origin. Case V(b) has local minima at origin and two  maximas at two symmetrically opposite sides of origin at \textit{non-zero p}. Both cases V(a) and V(b) are asymptotically vanishing.}\label{fig_case5}
	\end{center}
\end{figure}

\begin{figure}[h!] 
	\begin{center}
		\includegraphics[width=18cm]{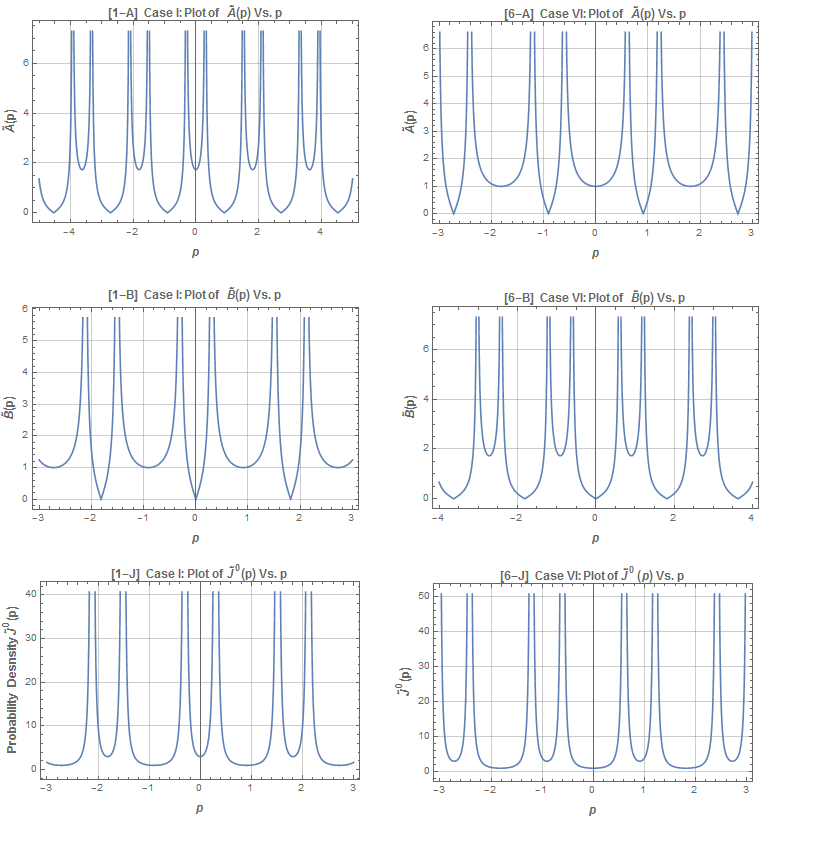}
		\caption{\textbf{Case I and Case VI}. The left column shows plots for Case 1 with $w=-2$. The right column shows plots for Case 6 with $w= +2$. Both the cases have unphysical solutions.
			}\label{fig_case1_6}
	\end{center}
\end{figure}

\begin{figure}[h!] 
	\begin{center}
		\includegraphics[width=18cm]{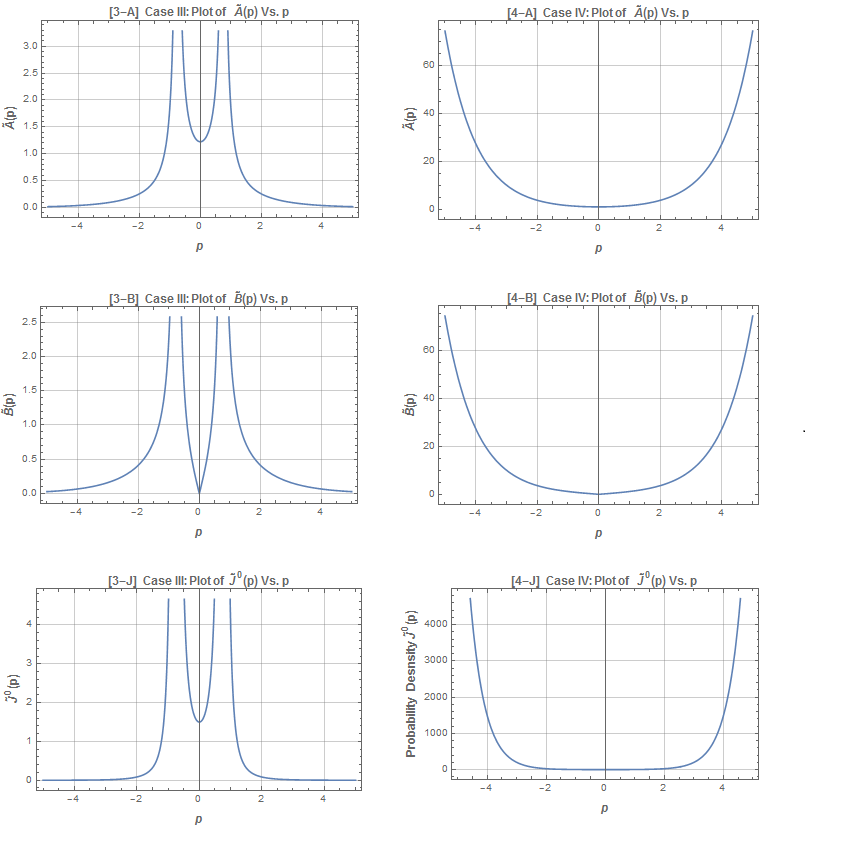}
		\caption{\textbf{Case III and Case IV}. Case III on the left, with $w=-0.5$. Case IV on the right, with $w=0$. Both the cases have unphysical solutions.
		}\label{fig_case3_4}
	\end{center}
\end{figure}
\newpage\cleardoublepage


\subsection{Attempting plane wave solutions}\label{sec:plane wave sol}
	
	For previous work on plane wave solutions of the non-linear Dirac equation see \cite{zecca,connor}. Our work in
	this section provides a more detailed analysis. We begin by considering the following plane wave ansatz:
	\begin{equation}
	\begin{bmatrix}
	F_1 \\ F_2 \\ G_1 \\ G_2 
	\end{bmatrix} = 
	\begin{bmatrix}
	u^0 \\ u^1 \\ \bar{v}_{0'} \\ \bar{v}_{1'}
	\end{bmatrix} e^{ik.x}
	\end{equation}
	
	With this ansatz, $\xi$ and $\xi^*$ are as follows:
	
	\begin{align}
	\xi &= u^A \bar{v}_{A'} \\
	\xi^* &= \bar{u}^{A'} v_A
	\end{align}
	
	Substituting the above ansatz in (\ref{eq:HD_in_cartesian_minkowski1} - \ref{eq:HD_in_cartesian_minkowski4}), we obtain the following equations$\colon$
	
	\begin{align}
	(k_0+k_3) u^0 + (k_1+ik_2) u^1 - \mu(\xi) \bar{v}_{0'}&= 0\\
	(k_0-k_3) u^1 + (k_1-ik_2)u^0  -\mu(\xi) \bar{v}_{1'}&= 0\\
	(k_0+k_3) \bar{v}_{1'} - (k_1-ik_2) \bar{v}_{0'} - \mu(\xi) u^1 &= 0\\
	(k_0-k_3) \bar{v}_{0'} - (k_1+ik_2) \bar{v}_{1'} - \mu(\xi) u^0 &= 0
	\end{align}
	
	Here  $\mu\big(\xi\big)=\sqrt{2}\big[b\big(l_2 \big)+ a\big(l_1)\xi \big]$ remains an undetermined quantity until a complete solution is obtained since $\xi$ is a function of the spinor. However, if we assume that $\xi$ is a real constant, we essentially end up with the usual Dirac equation with a ``modified mass'' $\mu\big(\xi \big)$. The equations can then be cast in matrix form:

	\begin{equation}
	\begin{pmatrix}
	(k_0+k_3) & (k_1+ik_2) & -\mu(\xi) &  0 \\
	(k_1-ik_2) & (k_0-k_3) & 0 & -\mu(\xi) \\
	0 & -\mu(\xi) &  - (k_1-ik_2) & (k_0+k_3) \\
	-\mu(\xi) & 0 & (k_0-k_3) & - (k_1+ik_2)
	\end{pmatrix} 
	\begin{pmatrix}
	u^0 \\ u^1 \\  \bar{v}_{0'} \\ \bar{v}_{1'}
	\end{pmatrix} = \begin{pmatrix}
	0 \\ 0 \\ 0 \\ 0
	\end{pmatrix}
	\end{equation}
	
	We work in the rest frame, and set  $k_1 = k_2 = k_3 = 0$. The matrix equation then reduces to$\colon$
	
	\begin{equation}
	\begin{pmatrix}
	k_0& 0 & -\mu(\xi) &  0 \\
	0 & k_0 & 0 & -\mu(\xi) \\
	0 & -\mu(\xi) & 0 & k_0 \\
	-\mu(\xi) & 0 & k_0 & 0
	\end{pmatrix} 
	\begin{pmatrix}
	u^0 \\ u^1 \\  \bar{v}_{0'} \\ \bar{v}_{1'}
	\end{pmatrix} = \begin{pmatrix}
	0 \\ 0 \\ 0 \\ 0
	\end{pmatrix}
	\end{equation}
	
	For a solution to exist, we require a null determinant. In other words, 
	
	\begin{align*}
	\big(k_0^2 - \mu(\xi)^2\big)^2 = 0 & \Rightarrow
	k_0 = \pm \mu(\xi)
	\end{align*}
	
	\underline{Case I}: $k_0 = \mu(\xi)$

	The general solution is of the form:
	
	\begin{equation}
	\begin{pmatrix}
	F_1\\ F_2\\ G_1\\ G_2\\
	\end{pmatrix}=\frac{\alpha_1}{\sqrt{V}}\begin{pmatrix}
	0\\ 1\\ 0\\1\\
	\end{pmatrix}e^{i\mu(\xi) x_0} +  \frac{\beta_1}{\sqrt{V}}\begin{pmatrix}
	1 \\ 0 \\ 1 \\ 0\\
	\end{pmatrix}e^{i\mu(\xi) x_0}
	\end{equation}
	where $|\alpha_1|^2+|\beta_1|^2 = 1$, and $V=6\pi l_{0}^{3}$ is the volume of the box in which the theory lives.
	
	\indent Here, $\xi$ and $\mu$ are as follows:
	\begin{align}
	\xi &= \frac{|\alpha_2|^2+|\beta_2|^2}{V} = \frac{1}{V}\\
	\mu &= \sqrt{2}\bigg{(}b + \frac{a}{V}\bigg{)} = \bigg{(}\frac{1}{ 2l_2} + \frac{l_{1}^{2}}{l_{0}^{3}}\bigg{)}~~~~~~ \text{where} ~ l_0^3>2l_1^2l_2
	\end{align}
	
	$\xi$ is indeed a real constant, and hence our approach is correct. Further we recall that,
	
	\begin{equation}
	\Psi= \begin{pmatrix}
	\psi_L\\
	\psi_R
	\end{pmatrix}= \begin{pmatrix}
	P^A\\
	\bar{Q}_{B\prime}
	\end{pmatrix}=\begin{pmatrix}
	P^0\\
	P^1\\
	\bar{Q}_{0^\prime}\\
	\bar{Q}_{1^\prime}\\
	\end{pmatrix}=\begin{pmatrix}
	P^0\\
	P^1\\
	-\bar{Q}^{1^\prime}\\
	\bar{Q}^{0^\prime}\\
	\end{pmatrix}=\begin{pmatrix}
	F_1\\
	F_2\\
	-G_1\\
	-G_2
	\end{pmatrix}
	\end{equation}
	
	Therefore, the actual spinor is given by$\colon$
	
	\begin{equation}
	\Psi=\frac{\alpha_1}{\sqrt{V}}\begin{pmatrix}
	0\\ 1\\ 0\\-1\\
	\end{pmatrix}e^{i(\mu_-) x_0} +  \frac{\beta_1}{\sqrt{V}}\begin{pmatrix}
	1 \\ 0 \\ -1 \\ 0\\
	\end{pmatrix}e^{i(\mu_-) x_0}
	\end{equation}

	Here, we have redefined $\mu\big(\xi \big)=\mu_-$ since the solution look like the negative frequency solutions to the Dirac equation with a mass $\mu_-$. This modified mass `$\mu_-$' is always positive. Hence $k_0 =  \mu_-$ is always positive in this case.
	
	\underline{Case II}: $k_0 = -\mu(\xi)$\\
	
	In this case, the general solution is of the form:
	
	\begin{equation}
	\begin{pmatrix}
	F_1\\ F_2\\ G_1\\ G_2\\
	\end{pmatrix}=\frac{\alpha_2}{\sqrt{V}}\begin{pmatrix}
	0\\ -1\\ 0\\1\\
	\end{pmatrix}e^{-i\mu(\xi) x_0} +  \frac{\beta_2}{\sqrt{V}}\begin{pmatrix}
	-1 \\ 0 \\ 1 \\ 0\\
	\end{pmatrix}e^{-i\mu(\xi) x_0}
	\end{equation}
	
	where, $|\alpha_2|^2+|\beta_2|^2 = 1$ is the normalization condition\\
	$\xi$, $\mu$ and $\Psi$ are given by:\\
	
	\begin{align}
	\xi &= \frac{-|\alpha_2|^2-|\beta_2|^2}{V} = \frac{-1}{V}\\
	\mu &= \sqrt{2}\bigg{(}b - \frac{a}{V}\bigg{)}=\bigg{(}\frac{1}{ 2l_2} - \frac{l_{1}^{2}}{l_{0}^{3}}\bigg{)} ~~~~~~ \text{where} ~ l_0^3>2l_1^2l_2\\
	\Psi &=\frac{\alpha_2}{\sqrt{V}}\begin{pmatrix}
	0\\ 1\\ 0\\1\\
	\end{pmatrix}e^{-i\mu_+ x_0} +  \frac{\beta_2}{\sqrt{V}}\begin{pmatrix}
	1 \\ 0 \\ 1 \\ 0\\
	\end{pmatrix}e^{-i\mu_+ x_0}
	\end{align}
	
	Once again we define $\mu\big(\xi \big)=\mu_+$ since this spinor looks like the positive frequency solution to the Dirac equation with a mass $\mu_+$. This modified mass `$\mu_+$' is always positive. Hence $k_0 = - \mu_+$ is always negative in this case.\\

	By substituting the expressions for the suitable length scales in various theories \big($l_1=0, l_2=\lambda_c/2$ for a torsionless theory, $l_1=L_{pl}, l_2=\lambda_c/2$ for standard ECD, $l_1=l_2=L_{CS}$ for modified ECD \big), and setting the value of fundamental constants to 1, we obtain the following table for the expressions of $\mu_+$ and $\mu_-$ in various cases:
	
	\begin{center}
		
		\begin{tabular}{ |c|c|c|c| }
			\hline
			& No torsion & Standard ECD & Modified ECD  \\ 
			\hline
			$\mu_+$ & $m_{1,2}$ & $m_{1,2}-\frac{L_{pl}^{2}}{l_{0}^{3}}$ & $\frac{1}{2L_{CS}}-\frac{L_{CS}^{2}}{l_{0}^{3}}$ \\ 
			\hline
			$\mu_-$ & $m_{1,2}$ & $m_{1,2}+\frac{L_{pl}^{2}}{l_{0}^{3}}$ & $\frac{1}{2L_{CS}}+\frac{L_{CS}^{2}}{l_{0}^{3}}$ \\ 
			\hline
		\end{tabular}
	\end{center}
	
	Corresponding to each value of $L_{CS}$, there are two values of mass $m_1$ and $m_2$. For the theory with no torsion $\mu_+\big(l_1,l_2\big)=\mu_-\big(l_1,l_2 \big)$; this equality breaks down when torsion is introduced, but is restored as $l_0$ tends to infinity. Note also that while $|m_{1,2}-\mu_+|=|m_{1,2}-\mu_-|$ is independent of $m_{1,2}$ for standard ECD, this isn't the case for modified ECD.

	\begin{figure}[h!]
		\hfill\includegraphics[keepaspectratio,scale=0.45]{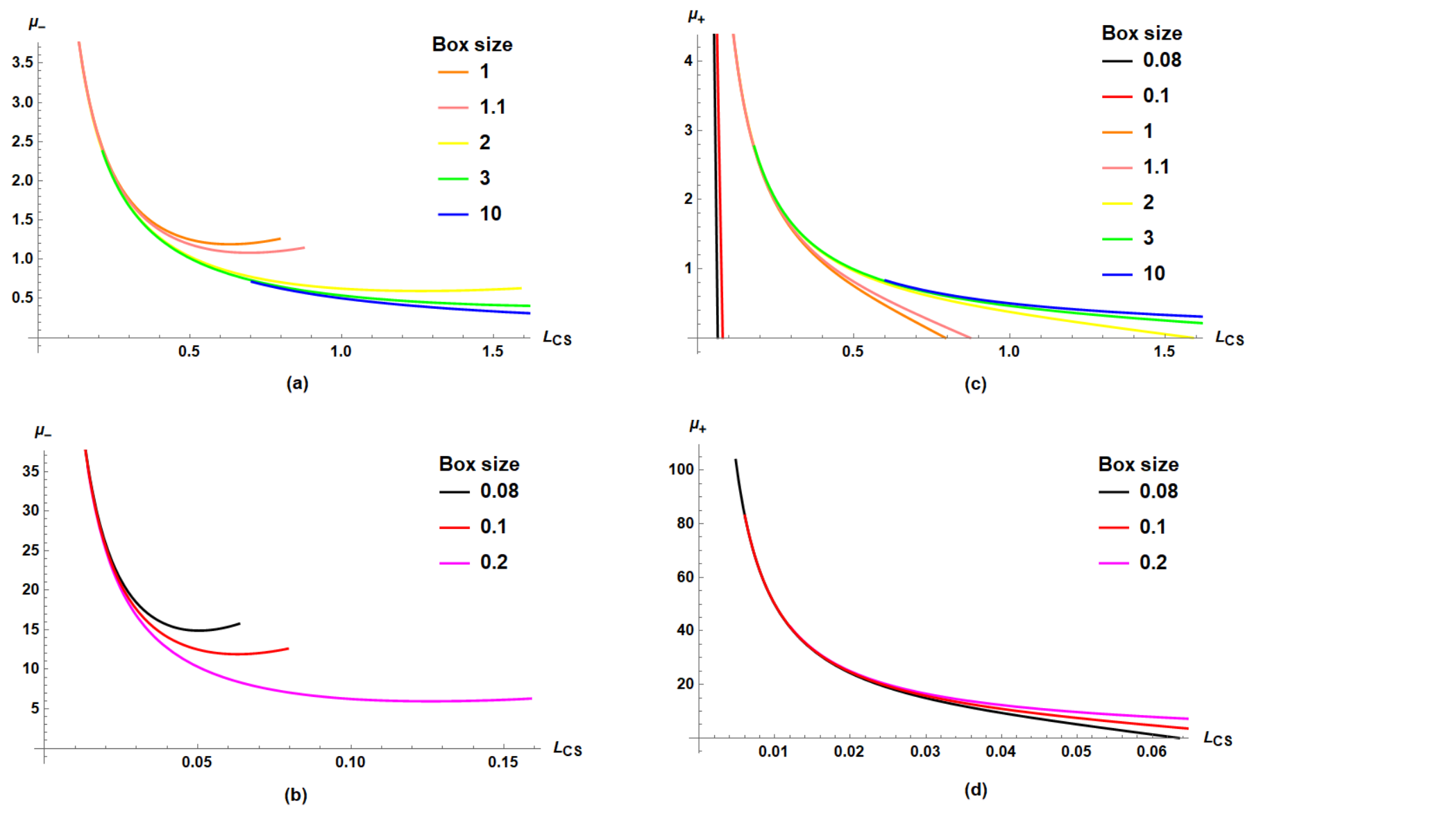}\hspace*{\fill}
		\caption{Plots for $\mu_-$ and $\mu_+$ as a function of $L_{CS}$ for various values of $l_0$  }
		\label{fig}
	\end{figure}

	Fig. \ref{fig} shows plots of $\mu_+$ and $\mu_-$ as a function of $L_{CS}$ (in the range $L_{pl}$ to $l_0$) for various values of $l_0$. Lengths are measured in units of $10^{23}$ $L_{pl}$. For a sense of scale, the $L_{CS}$ for an electron \big(and for its dual mass\big) is about $10^{22}L_{pl}=0.1$ in these units.\\
	
	 The symmetry between positive and negative frequency solutions is broken by torsion in a strange way. Further, the introduction of $L_{CS}$ introduces an interesting dependence of $\mu_+$ and $\mu_-$ on $L_{CS}$. In the standard ECD theory, $\mu_+$ ($\mu_-$) acquires a very small subtractive (additive)  ``correction term" which is proportional to $\frac{1}{l_{0}^{3}}$ and independent of the mass $m_{1,2}$. This term becomes insignificant as the box size becomes larger. 	But this situation changes dramatically for the modified $L_{CS}$ theory. $\mu_+$ decreases monotonically with $L_{CS}$ and increases monotonically with $l_0$. While $\mu_-$ decreases for $L_{CS} \leq l_0/4^{\frac{1}{3}}$, acquires a minimum at $L_{CS} = l_0/4^{\frac{1}{3}}$ and increases thereafter, it increases monotonically with $l_0$. The significance of the ``modified mass" $\mu$ in this case is still being investigated.

\subsection{Solution by reduction to (2+1) dim in cylindrical coordinates (t,r,$\phi$,z)}\label{sec:2+1}

After assuming $\partial_z = 0$, the HD equations in cylindrical coordinates [\ref{eq:HD_in_cylindrical_minkowski1} - \ref{eq:HD_in_cylindrical_minkowski4}] are as follows:
\begin{align}
r\partial_t F_1 + cr\partial_r F_2 e^{i\phi} + ic\partial_{\phi} F_2 e^{i\phi}  F_1 &= icr\sqrt{2} (b+a\xi)G_1 \\
r\partial_t F_2 + cr\partial_r F_1 e^{-i\phi} - ic\partial_{\phi} F_1 e^{-i\phi}&= icr\sqrt{2} (b+a\xi)G_2 \\
r\partial_t G_2 - cr\partial_r G_1 e^{-i\phi} + ic\partial_{\phi} G_1 e^{-i\phi} &= icr\sqrt{2} (b+a\xi^*)F_2 \\
r\partial_t G_1 - cr\partial_r G_2 e^{i\phi} - ic\partial_{\phi} G_2 e^{i\phi} &= icr\sqrt{2} (b+a\xi^*)F_1 
\end{align}
We now take the ansatz, $F_2 = G_2$ and $F_1 = -G_1$
\begin{align}
r\partial_t F_1 + r\partial_r F_2 e^{i\phi} + i\partial_{\phi} F_2 e^{i\phi} &= -ir\sqrt{2} (b+a\xi)F_1\\
r\partial_t F_2 + r\partial_r F_1 e^{-i\phi} - i\partial_{\phi} F_1 e^{-i\phi} &= ir\sqrt{2} (b+a\xi)F_2
\end{align}
We choose following ansatz in the above equation
\begin{align}
\begin{bmatrix}
F_1 \\ F_2
\end{bmatrix} =
\begin{bmatrix}
iA(r)e^{\frac{i\phi}{2}}\\ B(r)e^{\frac{-i\phi}{2}}
\end{bmatrix}e^{-i\omega t}
\end{align}
Putting this ansatz in above equations, we obtain the 2 differential equations as follows: 
\begin{align}
-rB\omega + r\partial_rA + \frac{A}{2} &= r\sqrt{2}[b+a(B^2-A^2)]B\\
rA\omega + r\partial_rB + \frac{B}{2} &= r\sqrt{2}[b+a(B^2-A^2)]A 
\end{align}
We add and subtract the two equations above and make the following substitution:
\begin{align}
\psi_1 &= B(r) + A(r)\\
\psi_2 &= B(r) - A(r)
\end{align}
so as to obtain
\begin{align}
-r\omega \psi_2 + r\psi_1' + \frac{\psi_1}{2} - r\sqrt{2}(b+a\psi_1\psi_2)\psi_1 &= 0\\
r\omega \psi_1 + r\psi_2' + \frac{\psi_2}{2} + r\sqrt{2}(b+a\psi_1\psi_2)\psi_2 &= 0
\end{align}
\indent With $\omega = 0$, we have the solutions: 
\begin{align}
\psi_1 = \bigg{[}\frac{c_2 e^{\sqrt{2}br}}{r^{\big{(}\frac{1-2\sqrt{2}ac_1}{2}\big{)}}} \bigg{]}~~~~~~~~~~~~~~
\psi_2 = \bigg{[} \frac{c_1 e^{-\sqrt{2}br} r^{\big{(}\frac{-1-2\sqrt{2}ac_1}{2}\big{)}}  }{c_2} \bigg{]}
\end{align}
\indent This is clearly unphysical because $\psi_1$ blows up $\forall$ non-zero $c_2$, and setting $c_2 = 0$ results in $\psi_2$ diverging. Thus, we conclude that a static solution to the above system of equation is unphysical, and $\omega$ cannot be zero. Further work to solve these equations numerically is in progress. 

\subsection{Solution by reduction to (3+1) Dim in spherical coordinates (t,r,$\theta$,$\phi$)}\label{sec:3+1}

We begin by putting following ansatz in HD equations with spherical coordinates:
\begin{align}
\begin{bmatrix}
F_1 \\ F_2 \\G_1 \\ G_2
\end{bmatrix} =
\begin{bmatrix}
R_{-\frac{1}{2}}(r) S_{-\frac{1}{2}}(\theta) e^{+i\phi/2}\\ R_{+\frac{1}{2}}(r)  S_{+\frac{1}{2}}(\theta)e^{-i\phi/2} \\ R_{+\frac{1}{2}}(r) S_{-\frac{1}{2}}(\theta) e^{+i\phi/2} \\ R_{-\frac{1}{2}}(r) S_{+\frac{1}{2}}(\theta)  e^{-i\phi/2}
\end{bmatrix}e^{-i\omega t}
\end{align}
With this ansatz, (\ref{eq:HD_in_SPC_minkowski1}) - (\ref{eq:HD_in_SPC_minkowski4}) become:
\begin{equation}
\begin{split}
\bigg{(}&-i\omega R_{-\frac{1}{2}} S_{-\frac{1}{2}} + \cos{\theta}R_{-\frac{1}{2}}' S_{-\frac{1}{2}}  - \frac{\sin{\theta}}{r}R_{-\frac{1}{2}} S_{-\frac{1}{2}}' + \frac{1}{2r\sin{\theta}}R_{+\frac{1}{2}} S_{+\frac{1}{2}} + \sin{\theta} R_{+\frac{1}{2}}'  S_{+\frac{1}{2}}   +\frac{\cos{\theta}}{r}R_{+\frac{1}{2}}  S_{+\frac{1}{2}}'\bigg{)} \\ &= i\sqrt{2}(b+a\xi)R_{+\frac{1}{2}} S_{-\frac{1}{2}}
\end{split}
\end{equation}
\begin{equation}
\begin{split}
\bigg{(}&-i\omega  R_{+\frac{1}{2}}  S_{+\frac{1}{2}} -\cos{\theta} R_{+\frac{1}{2}}'S_{+\frac{1}{2}} + \frac{\sin{\theta}}{r}R_{+\frac{1}{2}} S_{+\frac{1}{2}}' - \frac{1}{2r\sin{\theta}}R_{-\frac{1}{2}}  S_{-\frac{1}{2}}+ \sin{\theta} R_{-\frac{1}{2}}'  S_{-\frac{1}{2}} +\frac{\cos{\theta}}{r}R_{-\frac{1}{2}}  S_{-\frac{1}{2}}\bigg{)}' \\&= i\sqrt{2}(b+a\xi)R_{-\frac{1}{2}}(r) S_{+\frac{1}{2}}(\theta)
\end{split}
\end{equation}
\begin{equation}
\begin{split}
\bigg{(}&-i\omega R_{-\frac{1}{2}} S_{+\frac{1}{2}}+ \cos{\theta} R_{-\frac{1}{2}}' S_{+\frac{1}{2}} - \frac{\sin{\theta}}{r}R_{-\frac{1}{2}} S_{+\frac{1}{2}}'  + \frac{1}{2r\sin{\theta}}R_{+\frac{1}{2}} S_{-\frac{1}{2}} - \sin{\theta} R_{+\frac{1}{2}}' S_{-\frac{1}{2}} -\frac{\cos{\theta}}{r}R_{+\frac{1}{2}} S_{-\frac{1}{2}}' \bigg{)} \\ &= i\sqrt{2}(b+a\xi^*)R_{+\frac{1}{2}}(r) S_{+\frac{1}{2}}(\theta)
\end{split}
\end{equation}
\begin{equation}
\begin{split}
\bigg{(}&-i\omega R_{+\frac{1}{2}}(r) S_{-\frac{1}{2}}(\theta)-\cos{\theta} R_{+\frac{1}{2}}' S_{-\frac{1}{2}} + \frac{\sin{\theta}}{r}R_{+\frac{1}{2}} S_{-\frac{1}{2}}' -\frac{1}{2r\sin{\theta}}R_{-\frac{1}{2}} S_{+\frac{1}{2}} - \sin{\theta} R_{-\frac{1}{2}}' S_{+\frac{1}{2}} -\frac{ \cos{\theta}}{r}R_{-\frac{1}{2}}S_{+\frac{1}{2}}' \bigg{)} \\&= i\sqrt{2}(b+a\xi^*)R_{-\frac{1}{2}}S_{-\frac{1}{2}}
\end{split}
\end{equation}
where: 
\begin{align}
\xi &= R_{-\frac{1}{2}} S_{-\frac{1}{2}} \bar{R}_{+\frac{1}{2}} \bar{S}_{-\frac{1}{2}} + 
R_{+\frac{1}{2}}S_{+\frac{1}{2}} \bar{R}_{-\frac{1}{2}} \bar{S}_{-\frac{1}{2}} \\
\xi^* &= \bar{R}_{-\frac{1}{2}} \bar{S}_{-\frac{1}{2}} R_{+\frac{1}{2}} S_{-\frac{1}{2}} + 
\bar{R}_{+\frac{1}{2}} \bar{S}_{+\frac{1}{2}} R_{-\frac{1}{2}} S_{-\frac{1}{2}} 
\end{align}
Further work is in progress to investigate if this system of equations admits solitonic solutions.
\section{Summary}

In this paper, we formulated ECD theory in the NP formalism. The Dirac equation is carried to $U_4$ and presented (in NP) in (\ref{1-DEonU_4final} - \ref{4-DEonU_4final}). We have also provided a prescription for finding the covariant derivative on $U_4$ in NP formalism, thereby allowing one to calculate objects like the generic EM tensor on $U_4$ etc. We have calculated the spin density term which acts as a correction to the metric EM tensor; the two of which contribute together to the Einstein tensor (made up of Christoffel connections). In addition, the NP variables for the contorsion spin coefficients are also expressed in terms of the Dirac state. 

Solutions to linear Dirac equation on Minkowski space have been studied extensively. In this work, we attempted finding solutions to HD equations on Minkowski space with torsion. We explored whether presence of torsion induces any non-trivial (and physically relevant) modifications to the solutions for linear (non-torsional) case.  Solutions after reducing the problem to (1+1) dimension in the variables $(t,z)$ were found. We found a finite parameter range w $\in$ (0,1), where this solution vanishes at infinity in the non-static case and has finite maxima (or finite local minima) at origin. For   $w \in$ (1/3,1), the solution (and the probability density) decreases monotonically from a finite value at center and asymptotically reaches zero at infinity. This is the sought after finite solution - the `blob'. 

Plane wave solutions were found in section (\ref{sec:plane wave sol}). Next, we attempted finding solutions by reducing the problem to (2+1) dimensions in cylindrical coordinates with variables $(t,r,\phi)$. Static solutions to this were also found to be unphysical. However, finding non-static solutions to (2+1) case (given in section \ref{sec:2+1}) and the (3+1) case (given in section \ref{sec:3+1})  is work under progress. In future work we also hope to extend this investigation to Poincar\'e gauge gravity with propagating torsion. One of the principal goals of these studies is to look for torsion-induced nonsingular solitonic solutions of the non-linear Dirac equation.

\bigskip
\section{Appendices}
\subsection{Appendix A: Contorsion tensor ($K^{\mu\nu\alpha}$) components}\label{AppendixA}
Our aim is to write the contorsion tensor ($K^{\mu\nu\alpha}$) in the NP formalism eventually in terms of spinor components, with the contorsion tensor given by:
\begin{equation}\label{eq:calculating contorsion1}
K^{\mu\nu\alpha} = - kS^{\mu\nu\alpha} = 2i\pi l ^2 \bar{\psi}\gamma^{[\mu} \gamma^{\nu} \gamma^{\alpha]} \psi
\end{equation} 
\indent Note, only four independent components of this tensor is excited by the Dirac field. Writing explicitly in the NP formalism, i.e., null tetrad basis, we have:
\begin{equation}\label{eq:calculating contorsion2}
K_{(i)(j)(k)} = e_{(i)\mu}e_{(j)\nu}e_{(k)\alpha} K ^{\mu\nu\alpha}
\end{equation} 
where $e_{(i)\mu} = (l_{\mu},n_{\mu},m_{\mu},\bar{m}_{\mu})$ for $i = 0,1,2,3$
First, we consider the product $\gamma^{\alpha}\gamma^{\beta}\gamma^{\mu}$, defined as follows:
\begin{equation}
\gamma^{\alpha}\gamma^{\beta}\gamma^{\mu} =\begin{pmatrix} 0 & (\tilde{\sigma}^{\alpha})^*(\sigma^{\beta})^*(\tilde{\sigma}^{\mu})^* \\ (\sigma^{\alpha})^*(\tilde{\sigma}^{\beta})^*(\sigma^{\mu})^* & 0 \end{pmatrix}
=2\sqrt{2}
\begin{pmatrix}
0_{2\times2} & K_{01} \\
K_{10} & 0_{2\times2}\\
\end{pmatrix}
\end{equation}
where, explicitly, expanding out the Van der Waarden symbols, we have:
\begin{align}
K_{01}=&
\begin{bmatrix}
+nln-n\bar{m}m-\bar{m}mn+\bar{m}nm & -nl\bar{m}+n\bar{m}l+\bar{m}m\bar{m}-\bar{m}nl \\
-mln+m\bar{m}m+lmn-lnm & +ml\bar{m}-m\bar{m}l-lm\bar{m}+lnl 
\end{bmatrix}^{\alpha\beta\mu} \\
K_{10}=&
\begin{bmatrix}
+lnl-l\bar{m}m-\bar{m}ml+\bar{m}lm & +ln\bar{m}-l\bar{m}n -\bar{m}m\bar{m}+\bar{m}ln \\
+mnl-m\bar{m}m-nml+nlm & +mn\bar{m}-m\bar{m}n-nm\bar{m}+nln
\end{bmatrix}^{\alpha\beta\mu}
\end{align}
\indent With the expression for $\gamma^{\alpha}\gamma^{\beta}\gamma^{\mu}$, we can now define the world components of K. Next, we use (\ref{eq:calculating contorsion2}) to calculate the contorsion spin coefficients\cite{jogia_Griffiths} in the NP (null tetrad) basis. An an example, the solution for $\rho_1$ is given as:
\begin{equation}
\rho_1 = -K_{(0)(2)(3)} = -l_{\mu}m_{\nu}\bar{m}_{\alpha} K^{\mu\nu\alpha} = -2i\pi l^2 [l_{\mu}m_{\nu}\bar{m}_{\alpha}] \bar{\psi}\gamma^{[\mu}\gamma^{\nu}\gamma^{\alpha]}\psi
\end{equation}
The only quantity giving a non-zero scalar product when contracted with $l_{\mu}m_{\nu}\bar{m}_{\alpha}$ is $n^{\mu}\bar{m}^{\nu}m^{\alpha}$ and corresponding permutations (given the definition of $\gamma^{[\mu}\gamma^{\nu}\gamma^{\alpha]}$), giving $l_{\mu}m_{\nu}\bar{m}_{\alpha}n^{\mu}\bar{m}^{\nu}m^{\alpha}=1$. Thus:
\begin{equation}
\begin{split}
[l_{\mu}m_{\nu}\bar{m}_{\alpha}] \bar{\psi}\gamma^{[\mu}\gamma^{\nu}\gamma^{\alpha]}\psi &=  
\frac{\sqrt{2}}{3}\bar{\psi} \Bigg(
\left[\begin{smallmatrix}
0&0&\textendash1&0 \\ 0&0&0&0 \\ 0&0&0&0 \\0&0&0&0 
\end{smallmatrix}\right]
- \left[\begin{smallmatrix}
0&0&1&0\\0&0&0&0 \\ 0&0&0&0 \\ 0&0&0&0
\end{smallmatrix}\right]
+ \left[\begin{smallmatrix}
0&0&\textendash1&0\\ 0&0&0&0 \\ 0&0&0&0 \\ 0&0&0&0
\end{smallmatrix}\right]
- \left[\begin{smallmatrix}
0&0&0&0 \\ 0&0&0&0 \\ 0&0&0&0 \\ 0&0&\textendash1&0
\end{smallmatrix}\right] 
+ \left[\begin{smallmatrix}
0&0&0&0 \\ 0&0&0&0 \\ 0&0&0&0 \\ 0&1&0&0
\end{smallmatrix}\right]
- \left[\begin{smallmatrix}
0&0&0&0 \\ 0&0&0&0 \\ 0&0&0&0 \\ 0&\textendash1&0&0
\end{smallmatrix}\right]
\Bigg) \nonumber \\
&= \frac{\sqrt{2}}{3}\begin{pmatrix}
Q_0 & Q_1 & \bar{P}^{0'} & \bar{P}^{1'}
\end{pmatrix} \begin{pmatrix}
0&0&\textendash3&0 \\ 0&0&0&0 \\ 0&0&0&0 \\ 0&3&0&0
\end{pmatrix}\begin{pmatrix}
P^0 \\ P^1  \\ \bar{Q}_{0'}  \\  \bar{Q}_{1'} 
\end{pmatrix} \\
&= \sqrt{2}(\bar{P}^{1'} P^1 - Q^1  \bar{Q}^{1'})
\end{split}
\end{equation}
\indent This gives the full expression for $\rho$ (redefining the spinor components as prescribed):
\begin{equation}
\rho = -K_{(0)(2)(3)} = -2\sqrt{2}i\pi l^2 [F_2\bar{F}_2 - G_1\bar{G}_1]
\end{equation}
and similarly for the other spin coefficients.

\subsection{Appendix B: The Dirac equation in $U_4$}\label{AppendixB}

The Dirac equation in $U_4$ (the \textit{Hehl-Datta} equation) is given, in matrix form, as:
\begin{align}
i\begin{pmatrix} 0 & (\tilde{\sigma}^{\mu})^* \\ (\sigma^{\mu})^* & 0 \end{pmatrix}\nabla_{\mu}\begin{pmatrix} P^A \\ \bar{Q}_{B'} \end{pmatrix} = \frac{1}{2\sqrt{2}l}\begin{pmatrix} P^A \\ \bar{Q}_{B'} \end{pmatrix}
\end{align}
\indent Rewriting as a pair of matrix equations:
\begin{align}
\begin{pmatrix} \sigma^{\mu}_{00'}  & \sigma^{\mu}_{10'} \\ \sigma^{\mu}_{01'} & \sigma^{\mu}_{11'} \end{pmatrix}\nabla_{\mu}\begin{pmatrix} P^0 \\ P^1 \end{pmatrix} + \frac{i}{2\sqrt{2}l}\begin{pmatrix} -\bar{Q}^{1'}\\ \bar{Q}^{0'} \end{pmatrix} &= 0 \\
\begin{pmatrix} \sigma^{\mu}_{11'}  & -\sigma^{\mu}_{10'} \\ -\sigma^{\mu}_{01'} & \sigma^{\mu}_{00'} \end{pmatrix}\nabla_{\mu}\begin{pmatrix} -\bar{Q}^{1'}\\ \bar{Q}^{0'} \end{pmatrix} + \frac{i}{2\sqrt{2}l}\begin{pmatrix} P^0\\ P^1 \end{pmatrix} &= 0
\end{align}
\indent We will proceed to work through a solution for the first and third equation generated by this pair; the second and fourth follow along similar lines.

\textit{Equation 1}:

\begin{equation}
\begin{split}
\frac{i}{2\sqrt{2} l}\bar{Q}^{1'} &= \sigma^{\mu}_{00'}\nabla_{\mu} P^0 + \sigma^{\mu}_{10'}\nabla_{\mu}P^1\\
&= (\partial_{00'}P^0 + \Gamma^0_{~i00'}P^i) + (\partial_{10'}P^1 + \Gamma^1_{~i10'}P^i) \\
&= (D+\Gamma^0_{~000'}P^0 + \Gamma^0_{~100'}P^1) + (\delta^* + \Gamma^1_{~010'}P^0+ \Gamma^1_{~110'}P^1)  \\
&= (D+\Gamma_{1000'} - \Gamma_{0010'})P^0 + (\delta^* + \Gamma_{1100'} - \Gamma_{0110'})P^1 \\
&= (D + \epsilon^o + \epsilon_1 -\rho^o -\rho_1)P^0 + (\delta^* + \pi^o + \pi_1 -\alpha^o -\alpha_1)P^1  \\
&= (D+\epsilon_0 - \rho_0)P^0 + (\delta^{*} + \pi_0-\alpha_0)P^1 + \frac{3}{2}(\pi_1P^1 - \rho_1P^0)
\end{split}
\end{equation}
\newline
\indent\textit{Equation 3}:

\begin{equation}
\begin{split}
\frac{i}{2\sqrt{2} l}P^0 &= -\sigma^{\mu}_{11'}\nabla_{\mu} \bar{Q}^{1'} - \sigma^{\mu}_{10'}\nabla_{\mu}\bar{Q}^{0'} + \frac{i}{2\sqrt{2} l} P^0\\
&= -\bar{\sigma}^{\mu}_{11'}\nabla_{\mu} \bar{Q}^{1'} - \bar{\sigma}^{\mu}_{0'1}\nabla_{\mu}\bar{Q}^{0'} + \frac{i}{2\sqrt{2} l} P^0 \\
&=(\partial_{11'}\bar{Q}^{1'} + \bar{\Gamma}^{1'}_{~i'1'1}\bar{Q}^{i'}) + (\partial_{10'}\bar{Q}^{0'} + \bar{\Gamma}^{0'}_{~i'0'1}\bar{Q}^{i'}) \\
&=(\Delta\bar{Q}^{1'} + \bar{\Gamma}^{1'}_{~0'1'1}\bar{Q}^{0'} + \bar{\Gamma}^{1'}_{~1'1'1}\bar{Q}^{1'}) + (\delta^*\bar{Q}^{0'} + \bar{\Gamma}^{0'}_{~0'0'1}\bar{Q}^{0'} + \bar{\Gamma}^{0'}_{~1'0'1}\bar{Q}^{1'})  \\
&=(\Delta + \bar{\Gamma}_{1'1'0'1} - \bar{\Gamma}_{0'1'1'1}) \bar{Q}^{1'}+ (\delta^* + \bar{\Gamma}_{1'0'0'1} - \bar{\Gamma}_{0'0'1'1})\bar{Q}^{0'} \\
&=(\Delta + \mu^o+\mu_1- \gamma^o -\gamma_1) \bar{Q}^{1'}+ (\delta^* + \beta^o+\beta_1 - \tau^o - \tau_1)\bar{Q}^{0'}\\
&=(\Delta + \mu_0^* - \gamma_0^*)\bar{Q}^{1'} - (\delta^* +\beta_0^* -\tau_0^*)\bar{Q}^{0'} -
\frac{3}{2}(\mu_1 \bar{Q}^{1'} - \pi_1 \bar{Q}^{0'}) 
\end{split}
\end{equation}
where we have used the gamma matrices as defined in (\ref{gammamatrices}), computed the covariant derivatives using (\ref{eq:CD-Spinor}), (\ref{eq:CD-CoSpinor}) and the spin connections in terms of contorsion spin coefficients as given in (\ref{tab:Dirac spin coefficients U4}). Using this procedure, the four Dirac equations in $U_4$ are obtained as:
\begin{align}
(D+\epsilon_0 - \rho_0)F_1 + (\delta^{*} + \pi_0-\alpha_0)F_2 + \frac{3}{2}(\pi_1F_2 - \rho_1F_1) &= i b(l)G_1 \label{eq:Dirac eq U4 NP1.1}\\
(\Delta + \mu_0 - \gamma_0)F_2 + (\delta +\beta_0 -\tau_0)F_1 +
\frac{3}{2}(\mu_1 F_2 - \tau_1 F_1) &= i b(l)G_2 \\
(D+\epsilon_0^* - \rho_0^*)G_2 - (\delta + \pi_0^*-\alpha_0^*)G_1 - \frac{3}{2}(\tau_1G_1 - \rho_1 G_2 ) &= i b(l)F_2\\
(\Delta + \mu_0^* - \gamma_0^*)G_1 - (\delta^* +\beta_0^* -\tau_0^*)G_2 -
\frac{3}{2}(\mu_1 G_1 - \pi_1 G_2) &=i b(l)F_1 \label{eq:Dirac eq U4 NP1.4}
\end{align}
where we have also redefined $\{P,Q\} \rightarrow \{F,G\}$, as per the substitution in (\ref{diracspinor}) and to obtain a form that can be consistently compared with the primary source material in \cite{Chandru} (eqn. 108).
\subsection{Appendix C: Calculating $(T-S)_{\mu\nu}$}\label{AppendixC}

In theories which consider a balance between the Riemannian and torsional curvatures (such as in \cite{GRFessay2018}, the tensor $(T-S)_{\mu\nu}$ is of paramount importance. Vanishing $(T-S)_{\mu\nu}$ would take the form of a `balance condition', and represent a space with nonzero Riemannian curvature and torsion, but where the two exactly cancel each other out. The $(T-S)_{\mu\nu}$ tensor is defined as:
\begin{equation}
(T-S)_{\mu\nu} = T_{\mu\nu} - \frac{4\pi  l^{2} }{\hbar c} \eta_{\mu\nu}S^{\alpha\beta\lambda}S_{\alpha\beta\lambda}
\end{equation}
This tensor has 10 components. The 6 off-diagonal components are as follows:
\small
\begin{equation}
\begin{split}
(T-S)_{10}= \frac{i\hbar c}{4} \bigg{(}&
\bar{F}_1\partial_1 F_1 + \bar{F}_2 \partial_1 F_2 + \bar{G}_1\partial_1 G_1 + \bar{G}_2\partial_1 G_2 -\bar{F}_2 \partial_0 F_1 - \bar{F}_1 \partial_0 F_2 + \bar{G}_2 \partial_0G_1 + \bar{G}_1\partial_0 G_2 \\& -\partial_1\bar{F}_1 F_1 - \partial_1\bar{F}_2 F_2 -\partial_1\bar{G}_1 G_1 - \partial_1\bar{G}_2 G_2 
+\partial_0\bar{F}_2  F_1 + \partial_0\bar{F}_1  F_2 - \partial_0\bar{G}_2 G_1 - \partial_0\bar{G}_1 G_2 \bigg{)}
\end{split}
\end{equation}
\begin{equation}
\begin{split}
(T-S)_{20}= \frac{i\hbar c}{4 } \bigg{(}& \bar{F}_1\partial_2 F_1 + \bar{F}_2 \partial_2 F_2 + \bar{G}_1\partial_2 G_1 + \bar{G}_2\partial_2 G_2  + i\bar{F}_2\partial_0 F_1 - i \bar{F}_1\partial_0 F_2 - i\bar{G}_2 \partial_0G_1 + i \bar{G}_1\partial_0 G_2 \\& -\partial_2\bar{F}_1 F_1 -\partial_2 \bar{F}_2 F_2 - G_1\partial_2 \bar{G}_1 -\partial_2 \bar{G}_2 G_2  -i\partial_0\bar{F}_2 F_1 + i\partial_0 \bar{F}_1 F_2 +i\partial_0 \bar{G}_2 G_1 -i\partial_0  \bar{G}_1 G_2  \bigg{)}
\end{split}
\end{equation}
\begin{equation}
\begin{split}
(T-S)_{30}= \frac{i\hbar c}{4 } \bigg{(}& \bar{F}_1\partial_3 F_1 + \bar{F}_2 \partial_3F_2 + \bar{G}_1\partial_3 G_1 + \bar{G}_2\partial_3G_2  -\bar{F}_1 \partial_0 F_1 + \bar{F}_2\partial_0 F_2 + \bar{G}_1 \partial_0G_1 - \bar{G}_2\partial_0 G_2 \\ & -\partial_3\bar{F}_1 F_1 - \partial_3\bar{F}_2 F_2 -\partial_3 \bar{G}_1 G_1 -\partial_3\bar{G}_2 G_2 +\partial_0\bar{F}_1 F_1 -\partial_0\bar{F}_2 F_2 -\partial_0 \bar{G}_1 G_1 +\partial_0 \bar{G}_2 G_2 \bigg{)}
\end{split}
\end{equation}
\begin{equation}
\begin{split}
(T-S)_{21} = \frac{i\hbar c}{4} \bigg{(}& i\bar{F}_2 \partial_1 F_1 - i \bar{F}_1\partial_1 F_2 - i \bar{G}_2\partial_1 G_1 + i \bar{G}_1\partial_1 G_2  -\bar{F}_2 \partial_2 F_1 - \bar{F}_1\partial_2 F_2 + \bar{G}_2 \partial_2 G_1 + \bar{G}_1\partial_2 G_2 \\ & -i\partial_1\bar{F}_2 F_1 + i\partial_1 \bar{F}_1 F_2 + i\partial_1  \bar{G}_2 G_1 -i\partial_1  \bar{G}_1 G_2 + \partial_2\bar{F}_2F_1 + \partial_2 \bar{F}_1 F_2 -\partial_2\bar{G}_2 G_1 -\partial_2\bar{G}_1 G_2 \bigg{)}
\end{split}
\end{equation}
\begin{equation}
\begin{split}
(T-S)_{31} = \frac{i\hbar c}{4} \bigg{(}& -\bar{F}_1 \partial_1 F_1 + \bar{F}_2\partial_1 F_2 + \bar{G}_1\partial_1 G_1 - \bar{G}_2 \partial_1 G_2\  -\bar{F}_2\partial_3 F_1 -\bar{F}_1 \partial_3F_2 +\bar{G}_2\partial_3 G_1 + \bar{G}_1\partial_3 G_2 \\& +\partial_1\bar{F}_1F_1 -\partial_1 \bar{F}_2F_2 -\partial_1 \bar{G}_1G_1 +\partial_1\bar{G}_2 G_2 +\partial_3 \bar{F}_2F_1 + \partial_3 \bar{F}_1 F_2 -\partial_3 \bar{G}_2 G_1 -\partial_3\bar{G}_1G_2	\bigg{)}
\end{split}
\end{equation}
\begin{equation}
\begin{split}
(T-S)_{32}(\{\}) = \frac{i\hbar c}{4} \bigg{(}& -\bar{F}_1 \partial_2  F_1 + \bar{F}_2 \partial_2 F_2 + \bar{G}_1 \partial_2 G_1 - \bar{G}_2\partial_2  G_2  + i \bar{F}_2\partial_3 F_1 - i\bar{F}_1\partial_3 F_2 - i \bar{G}_2\partial_3G_1 + i \bar{G}_1\partial_3 G_2  \\&  +\partial_2 \bar{F}_1 F_1 -\partial_2 \bar{F}_2 F_2 -\partial_2 \bar{G}_1 G_1 +\partial_2  \bar{G}_2 G_2  -i\partial_3\bar{F}_2 F_1 + i\partial_3\bar{F}_1F_2 + i\partial_3 \bar{G}_2G_1 - i\partial_3\bar{G}_1 G_2  \bigg{)}
\end{split}
\end{equation}
\normalsize
The diagonal components are as follows:
\small
\begin{equation}
\begin{split}
(T-S)_{00} &= \frac{i\hbar c}{2} \bigg{(} \bar{G}_1\partial_0G_1 + \bar{G}_2\partial_0G_2 - \partial_0\bar{G}_1G_1 - \partial_0\bar{G}_2 G_2
+ \bar{F}_1\partial_0F_1 + \bar{F}_2\partial_0F_2 - \partial_0\bar{F}_1 F_1 - \partial_0\bar{F}_2F_2\bigg{)} - 6\pi \hbar c l^2\xi\xi^*
\end{split}
\end{equation}
\begin{equation}
\begin{split}
(T-S)_{11} &= \frac{i\hbar c}{2} \bigg{(}-\bar{F}_2\partial_1 F_1 - \bar{F}_1\partial_1 F_2 + \bar{G}_2\partial_1 G_1 + \bar{G}_1\partial_1 G_2  +\partial_1\bar{F}_2F_1 +\partial_1 \bar{F}_1 F_2 -\partial_1\bar{G}_2 G_1 - \partial_1\bar{G}_1 G_2 \bigg{)}+ 6\pi \hbar c l^2\xi\xi^*
\end{split}
\end{equation}
\begin{equation}
\begin{split}
(T-S)_{22} &= \frac{i\hbar c}{2} \bigg{(}  i\bar{F}_2\partial_2 F_1 - i\bar{F}_1 \partial_2 F_2 - i\bar{G}_2\partial_2  G_1 + i\bar{G}_1 \partial_2 G_2
- i\partial_2 \bar{F}_2 F_1 + i\partial_2  \bar{F}_1 F_2 +i\partial_2  \bar{G}_2 G_1 - i\partial_2 \bar{G}_1 G_2	\bigg{)}+ 6\pi \hbar c l^2\xi\xi^*
\end{split}
\end{equation}
\begin{equation}
\begin{split}
(T-S)_{33} &= \frac{i\hbar c}{2 } \bigg{(} -\bar{F}_1\partial_3 F_1 + \bar{F}_2 \partial_3F_2 + \bar{G}_1 \partial_3G_1 - \bar{G}_2 \partial_3G_2 +\partial_3\bar{F}_1 F_1 - \partial_3\bar{F}_2F_2 -\partial_3\bar{G}_1 G_1 +\partial_3 \bar{G}_2 G_2\bigg{)} + 6\pi \hbar c l^2\xi\xi^*
\end{split}
\end{equation}
\normalsize
We can now calculate this tensor for the various solutions to the HD equations on Minkowski space with torsion, to probe the feasibility of a balance condition.  

\subsubsection*{$(T-S)_{\mu\nu}$ for non-static solutions in $1+1$ dim $(t,z)$}
\begin{equation}\label{eq:T-S_1+1_f(A,B)}
(T-S)_{\mu\nu} = \hbar c\begin{bmatrix}
\bigg{(}\Lambda[A^2+B^2] - \frac{a[A^2-B^2]^2}{2\sqrt{2}} \bigg{)}&  0  & -\Lambda AB  & 0 \\
0 & \bigg{(}\frac{a[A^2-B^2]^2}{2\sqrt{2}} \bigg{)} & 0  & 0  \\
-\Lambda AB & 0 & \bigg{(}\frac{a[A^2-B^2]^2}{2\sqrt{2}} \bigg{)}  & 0 \\
0 & 0 & 0 & \bigg{(}[AB' - BA'] + \frac{a[A^2-B^2]^2}{2\sqrt{2}} \bigg{)} 
\end{bmatrix}
\end{equation}

$\Lambda$ is a free parameter in the solution. We will analyze this tensor "T-S" for various types of values of $\Lambda$.

\subsection{Appendix D: The linear (torsionless) Dirac equation in $1+1$ dimensions}\label{AppendixD}

The vanishing of torsion is characterized by the limit $a(l_2) = 3\sqrt{2}\pi L_{Pl}^2 \longrightarrow 0$. So in a torsionless case, the differential equations become (with dimensionless constants):
\begin{align}
B' &= (1-w)A \\
A' &= (1+w)B  
\end{align}
Their solutions in various special cases are plotted below:

\begin{figure}[h!] 
	\begin{center}
		\includegraphics[width=18cm]{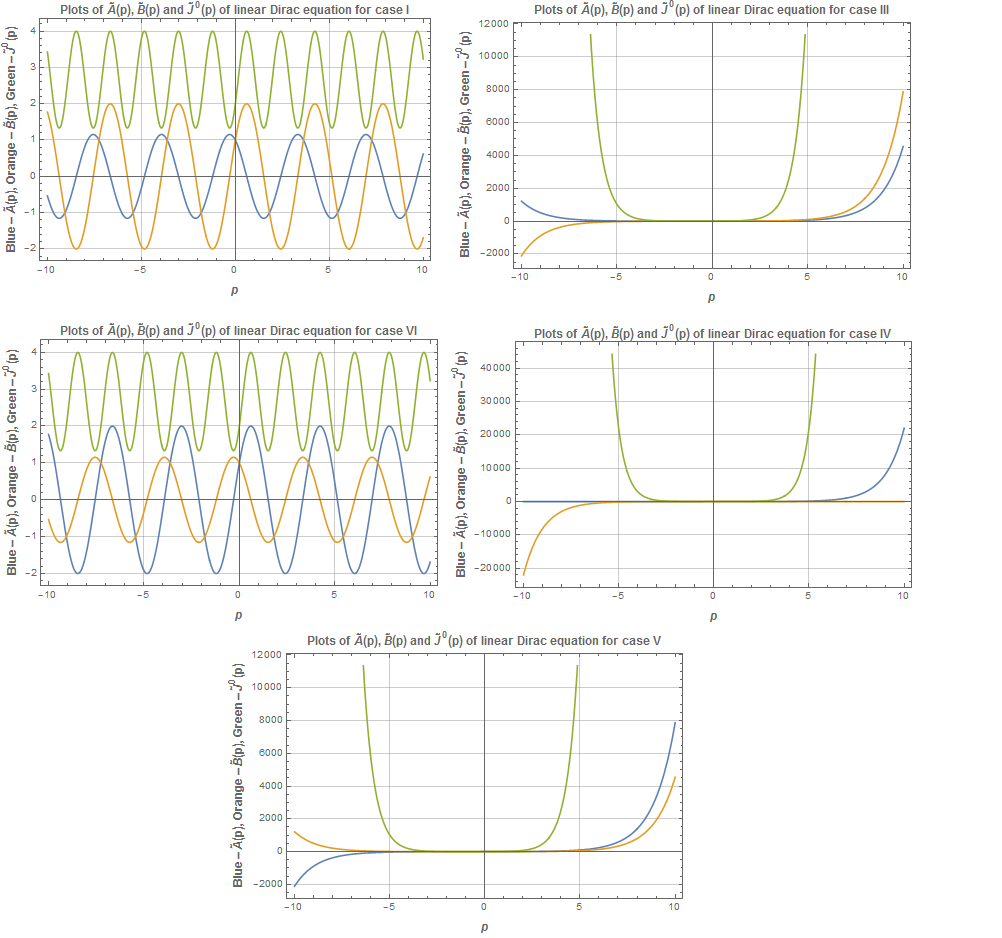}
		\caption{\textbf{Solutions to the linear (torsionless) Dirac equations}. Only the plane-wave solutions (Cases I, VI) are physical.
		}\label{fig:nontorsion Dirac case}
	\end{center}
\end{figure}

\end{document}